\documentclass{natureprintstyle}

\usepackage{graphicx,amsmath}
\usepackage{lscape}
\usepackage{color}
\usepackage{url}
\usepackage{ulem}
\usepackage{multirow}
\usepackage{caption}
\usepackage[]{graphicx}
\usepackage{amssymb}

\usepackage{astjnlabbrev-nature} 

\def\pasa{PASA}%

\title{The close environments of accreting massive black holes are shaped by radiative feedback}

\author{Claudio Ricci$^{1,2,3}$, Benny Trakhtenbrot$^{4}$, Michael J. Koss$^{4,5}$, Yoshihiro Ueda$^{6}$, Kevin Schawinski$^{4}$, Kyuseok Oh$^{4}$, Isabella Lamperti$^{4}$, Richard Mushotzky$^{7}$, Ezequiel Treister$^{1}$, Luis C. Ho$^{3,8}$, Anna Weigel$^{4}$, Franz E. Bauer$^{1,9,10}$, Stephane Paltani$^{11}$, Andrew C. Fabian$^{12}$, Yanxia Xie$^{3,8}$, Neil Gehrels$^{13}$}

\begin{document}

\maketitle

\begin{affiliations}
 \item Instituto de Astrof\'{\i}sica, Facultad de F\'{i}sica, Pontificia Universidad Cat\'{o}lica de Chile, Casilla 306, Santiago 22, Chile
 \item Chinese Academy of Sciences South America Center for Astronomy and China-Chile Joint Center for Astronomy, Camino El Observatorio 1515, Las Condes, Santiago, Chile
 \item Kavli Institute for Astronomy and Astrophysics, Peking University, Beijing 100871, China
 \item Institute for Astronomy, Department of Physics, ETH Zurich,Wolfgang-Pauli-Strasse 27, CH-8093 Zurich, Switzerland
 \item Eureka Scientific Inc., 2452 Delmer St. Suite 100, Oakland, CA 94602, USA
 \item Department of Astronomy, Kyoto University, Kyoto 606-8502, Japan
 \item Department of Astronomy and Joint Space-Science Institute, University of Maryland, College Park, MD 20742, USA
 \item Department of Astronomy, School of Physics, Peking University, Beijing 100871, China
  \item Space Science Institute, 4750 Walnut Street, Suite 205, Boulder, Colorado 80301, USA
  \item Millenium Institute of Astrophysics, Santiago, Chile
  \item Department of Astronomy, University of Geneva, ch. d'Ecogia 16, CH-1290 Versoix, Switzerland
  \item Institute of Astronomy, Madingley Road, Cambridge CB3 0HA, UK
  \item NASA Goddard Space Flight Center, Greenbelt, MD 20771, USA

\end{affiliations}

\begin{abstract}

The large majority of the accreting supermassive black holes in the Universe are obscured by large columns of gas and dust\cite{Burlon:2011dk,Ueda:2014ix,Ricci:2015tg}. 
The location and evolution of this obscuring material have been the subject of intense research in the past decades\cite{Elitzur:2006wq,Merloni:2014qv}, and are still highly debated. 
A decrease in the covering factor of the circumnuclear material with increasing accretion rates has been found by studies carried out across the electromagnetic spectrum\cite{Ueda:2003qf,Burlon:2011dk,Maiolino:2007ii,Treister:2008ff}. 
The origin of this trend has been suggested to be driven either 
by the increase in the inner radius of the obscuring material with incident luminosity due to the sublimation of dust\cite{Lawrence:1991vn}; 
by the gravitational potential of the black hole\cite{Lamastra:2006hl}; 
by radiative feedback\cite{Fabian:2006lq,Menci:2008fj,Fabian:2008hc,Fabian:2009ez}; 
or by the interplay between outflows and inflows\cite{Wada:2015oz}. 
However, the lack of a large, unbiased and complete sample of accreting black holes, with reliable information on gas column density, luminosity and mass, has left the main physical mechanism regulating obscuration unclear. 
Using a systematic multi-wavelength survey of hard X-ray-selected black holes, here we show that radiation pressure on dusty gas is indeed the main physical mechanism regulating the distribution of the circumnuclear material. 
Our results imply that the bulk of the obscuring dust and gas in these objects is located within the sphere of influence of the black hole (i.e., a few to tens of parsecs), and that it can be swept away even at low radiative output rates.
The main physical driver of the differences between obscured and unobscured accreting black holes is therefore their mass-normalized accretion rate.

\end{abstract}

\paragraph*{}
\label{sec:introduction}

\paragraph*{}
\label{sec:results}
Our group has carried out a large multi-wavelength study of the 836 accreting supermassive black holes (i.e., active galactic nuclei or AGN) detected by the all-sky hard X-ray (14--195\,keV) {\it Swift} Burst Alert Telescope survey\cite{Gehrels:2004kx,Barthelmy:2005uq} (see \S\ref{sec:sample} of the Methods). The energy range covered by {\it Swift}/BAT makes it ideal for studying the characteristics and evolution of the absorbing material surrounding the AGN, being unaffected by obscuration up to column densities $N_{\rm H}\simeq 10^{24}\rm\,cm^{-2}$. Moreover, this is a local sample of AGN, with a median redshift of $z\simeq0.037$ (distance $D\simeq162$\,Mpc), for which reliable estimates of the black hole mass ($M_{\rm BH}$) can be obtained. We analyzed spectroscopic data in the X-ray (0.3--150\,keV) and optical bands, which allowed us to place strong estimates on the column densities and intrinsic X-ray luminosities\cite{Ricci:2017aa}, and to infer the black hole mass for 392 AGN\cite{Koss:2017aa}. These measurements allow us to study the relation between obscuration and accretion properties for a large number of hard X-ray selected accreting black holes, in unprecedented detail.

\begin{figure}[t!]
 \centering
\includegraphics[width=0.48\textwidth]{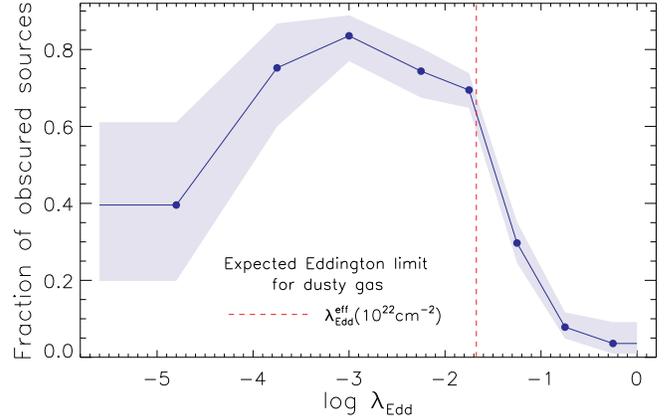}
 \caption{\textbf{Relation between the fraction of obscured AGN and the Eddington ratio.} The fraction of obscured Compton-thin [$10^{22}\leq (N_{\rm H}/\rm cm^{-2})< 10^{24}$] sources shown as a function of the Eddington ratio $\lambda_{\rm Edd}$ (i.e. the AGN luminosity normalized by the maximum value for solar-metalicity, fully-ionized, dust-free gas in a spherical geometry) for our hard X-ray selected sample in the $10^{-5.6} \leq \lambda_{\rm Edd}< 1$ range. The values are normalized to unity in the $10^{20} \leq (N_{\rm H}/\rm cm^{-2})< 10^{24}$ interval. The shaded area represents the 16th and 84th quantiles of a binomial distribution\cite{Cameron:2011cl}. 
The vertical red dashed line represents the effective Eddington limit for a dusty gas\cite{Fabian:2009ez} with $N_{\rm H}=10^{22}\rm cm^{-2}$ (see \S\ref{sec:EddNHdiagram}).  
The figure shows that the covering factor of the obscuring material with $10^{22} \leq (N_{\rm H}/\rm cm^{-2})< 10^{24}$ decreases sharply around the Eddington limit for dusty gas, highlighting the fact that radiation pressure strongly affects obscuration in AGN.}
\label{fig:FobsVsEddratio}
\end{figure}

\noindent In Figure\,\,\ref{fig:FobsVsEddratio} we show, for this large and unbiased sample of AGN, the relation between the fraction of Compton-thin obscured sources [$10^{22} \leq  (N_{\rm H}/\rm cm^{-2})< 10^{24}$] and the mass-normalized accretion rate (i.e., the Eddington ratio, $\lambda_{\rm Edd}\propto L/M_{\rm BH}$, see \S\ref{sec:EddNHdiagram}). The fraction of obscured sources, i.e. the covering factor of the Compton-thin circumnuclear material, exhibits a sharp decline at $\lambda_{\rm Edd}\simeq 0.02-0.05$ (dropping from $f_{\rm obs}\simeq80$\% to $\simeq30$\%). This range in $\lambda_{\rm Edd}$ corresponds to the Eddington limit for dusty gas\cite{Fabian:2008hc} (i.e., the effective Eddington limit, $\lambda_{\rm Edd}^{\rm eff}$). The cross-section of gas coupled with dust is in fact higher than that of ionized hydrogen\cite{Fabian:2006lq}, and the Eddington limit can be reached at $\lambda_{\rm Edd}=\lambda_{\rm Edd}^{\rm eff}\simeq 10^{-2}$ (\S\ref{sec:EddNHdiagram}). The values of $f_{\rm obs}$ are normalized over the $N_{\rm H}=10^{20}-10^{24}\rm\,cm^{-2}$ range, and the uncertainties reported on the fractions of unobscured and obscured Compton-thin AGN represent the 16th and 84th quantiles of a binomial distribution\cite{Cameron:2011cl}.

\begin{figure}[t!]
 \centering
\includegraphics[width=0.48\textwidth]{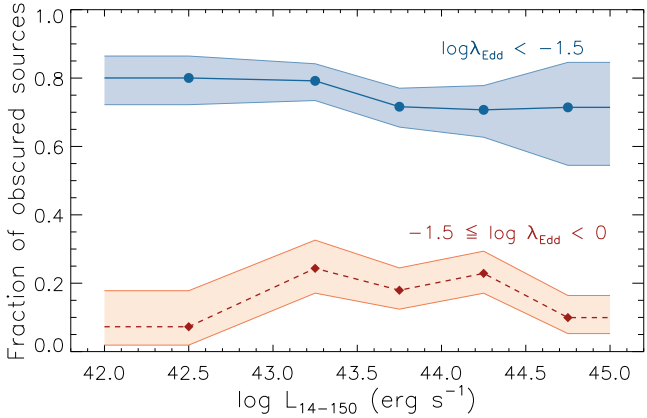}
 \caption{\textbf{Relation between the fraction of obscured AGN and the luminosity for different ranges of Eddington ratio.} The fraction of obscured Compton-thin [$22 \leq \log (N_{\rm H}/\rm cm^{-2})< 24$] sources shown as a function of the intrinsic (i.e. absorption-corrected) X-ray luminosity for the sources of our sample divided into two ranges of Eddington ratio: $\lambda_{\rm Edd}< 10^{-1.5}$ (blue) and $10^{-1.5} \leq  \lambda_{\rm Edd}< 1$ (red). The shaded area represents the 16th and 84th quantiles of a binomial distribution\cite{Cameron:2011cl}. The figure shows that any luminosity dependence of the covering factor of the obscuring material disappears when different intervals of the Eddington ratio are considered. This implies that the parameter driving the evolution of the circumnuclear material is the Eddington ratio and not the luminosity, and that the bulk of the obscuring material is located within the region in which the gravitational potential of the supermassive black hole dominates over that of the galaxy.
 }
 \label{fig:FobsVsLeddratiobins}
\end{figure}

\noindent The luminosity-dependence of the fraction of obscured sources, which is clearly observed for the whole sample (\S\ref{sect:obsLuminosity}), disappears almost entirely when dividing the sources into bins of $\lambda_{\rm Edd}$ (Figure\,\,\ref{fig:FobsVsLeddratiobins}). On the other hand, separating the sources in different ranges of luminosity and $M_{\rm BH}$ we find the same trend between $f_{\rm obs}$ and $\lambda_{\rm Edd}$ obtained for the whole sample (\S\ref{sect:obsEddratio}). 
Similar to what is observed for $f_{\rm obs}$,  the median column density decreases sharply from $ N_{\rm H}\simeq10^{23}\rm\,cm^{-2}$ to $N_{\rm H}\simeq10^{20.5}\rm\,cm^{-2}$ at $\lambda_{\rm Edd}\gtrsim 0.02$; the same behaviour is also found when dividing the sample to subsets covering different ranges of luminosities and black hole mass (\S\ref{sect:obsEddratio}).  These results show that the main physical mechanism regulating the covering factor and the amount of material around black holes is radiation pressure on dusty gas.

\begin{figure}[t!]
 \centering
\includegraphics[width=0.48\textwidth]{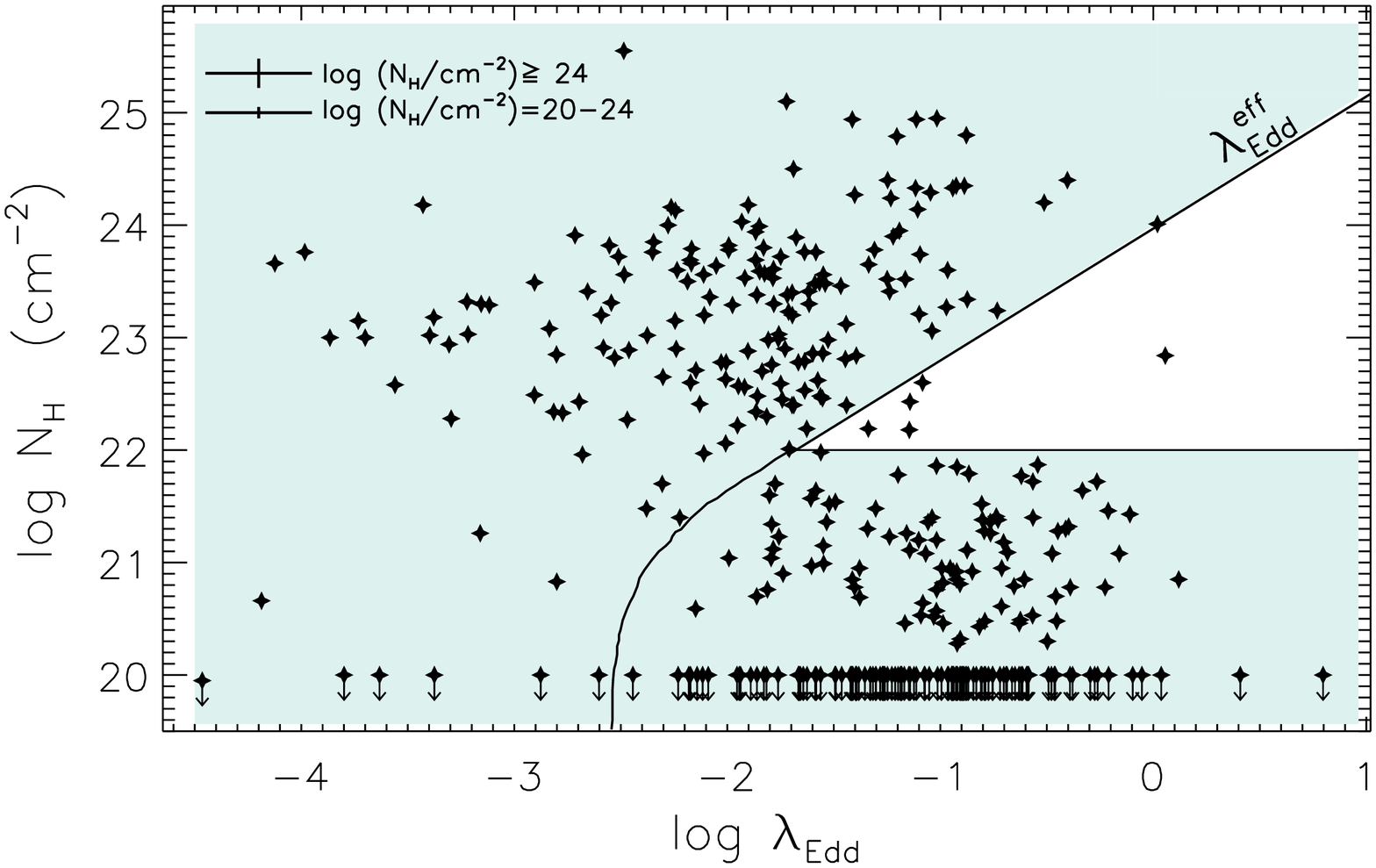}
 \caption{\textbf{Eddington ratio--column density diagram.}  The continuous line represents the effective Eddington limit for different values of column density, for a standard dust grain composition of the interstellar medium\cite{Fabian:2009ez}. The horizontal line at $N_{\rm H}=10^{22}\rm\,cm^{-2}$ shows the region where dust lanes from the host galaxy are typically expected to contribute to the line-of-sight column density. The \textit{blowout region}, i.e. the place in the $N_{\rm H}-\lambda_{\rm Edd}$ diagram where radiation pressure would push away the obscuring material [i.e. $\lambda_{\rm Edd}>\lambda_{\rm Edd}^{\rm eff}(N_{\rm H})$], is shown in white. The region where the Eddington ratio is below the Eddington limit for dusty gas is shown in green. The stars represent the objects of our sample; unobscured sources were assigned the upper limit of $N_{\rm H}=10^{20}\rm\,cm^{-2}$. The two crosses on the top left show the typical 1-$\sigma$ confidence intervals for Compton-thick (top) and Compton-thin (bottom) AGN.
 }
 \label{fig:EddratioNHdiag}
\end{figure}

\noindent The effective Eddington limit increases with $N_{\rm H}$, due to the larger amounts of material that radiation pressure needs to push away for larger column densities\cite{Fabian:2006lq}. In Figure\,\,\ref{fig:EddratioNHdiag} we illustrate the distribution of the sources in our sample in the $N_{\rm H}-\lambda_{\rm Edd}$ diagram\cite{Fabian:2008hc,Fabian:2009ez} (\S\ref{sec:EddNHdiagram}). The white area corresponds to values of $\lambda_{\rm Edd}$ above the effective Eddington limit for a given column density. AGN in this region are expected to expel their circumnuclear material through radiation pressure. Only $1.4^{+0.7}_{-0.5}\%$ of the sources in our sample are found in this blowout region, which implies that the radiative feedback from the AGN can very efficiently clear out the close environment of the accreting black hole at $\lambda_{\rm Edd} \ll 1$. For a given value of $\lambda_{\rm Edd}$ the range of column densities observed is likely related to different viewing angles, as foreseen by the standard unification model of AGN\cite{Antonucci:1993fu}, which ascribes most of the differences between obscured and unobscured AGN to different inclinations with respect to a toroidal structure of gas and dust that surrounds the accreting source.

\noindent We find a tentative decrease of $f_{\rm obs}$ at $\lambda_{\rm Edd}\leq 10^{-4}$, with only $40^{+21}_{-20}\%$ of the sources being obscured by Compton-thin material. A similar trend is found for the median column density. 
It has been proposed that the obscuring material is related to outflows produced by radiation pressure from the accretion disk\cite{Elitzur:2006wq}. One of the postulates of such a model is that at low $\lambda_{\rm Edd}$ only a limited amount of obscuring material is puffed up, leading to a smaller covering factor. This is similar to what is shown here. Alternatively, the low Eddington ratio itself could be driven by the limited amount of dusty gas available for the accretion process.

\paragraph*{}
\label{sec:discussion}

\begin{figure}[t!]
 \centering
\includegraphics[width=0.48\textwidth]{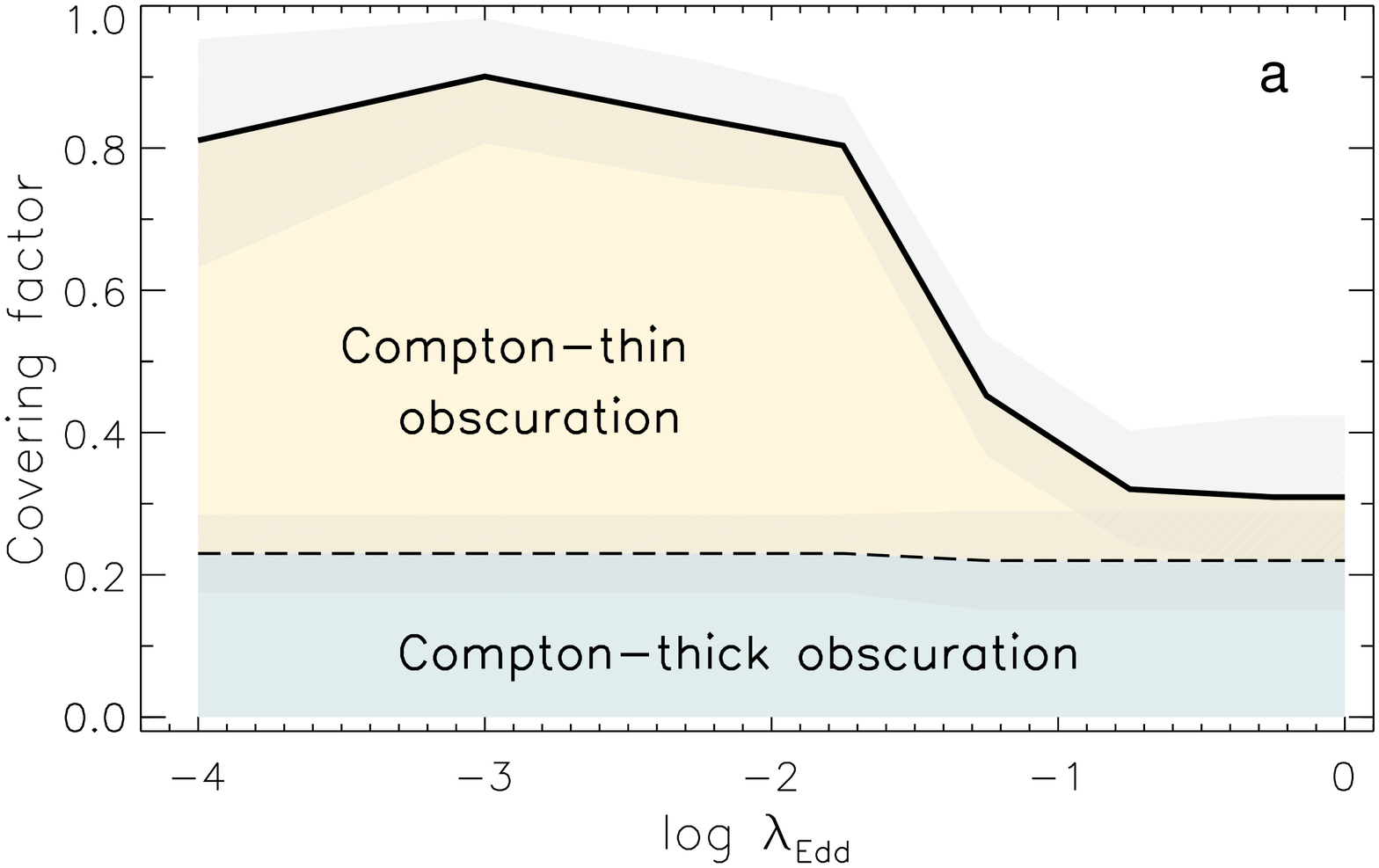}
\includegraphics[width=0.48\textwidth]{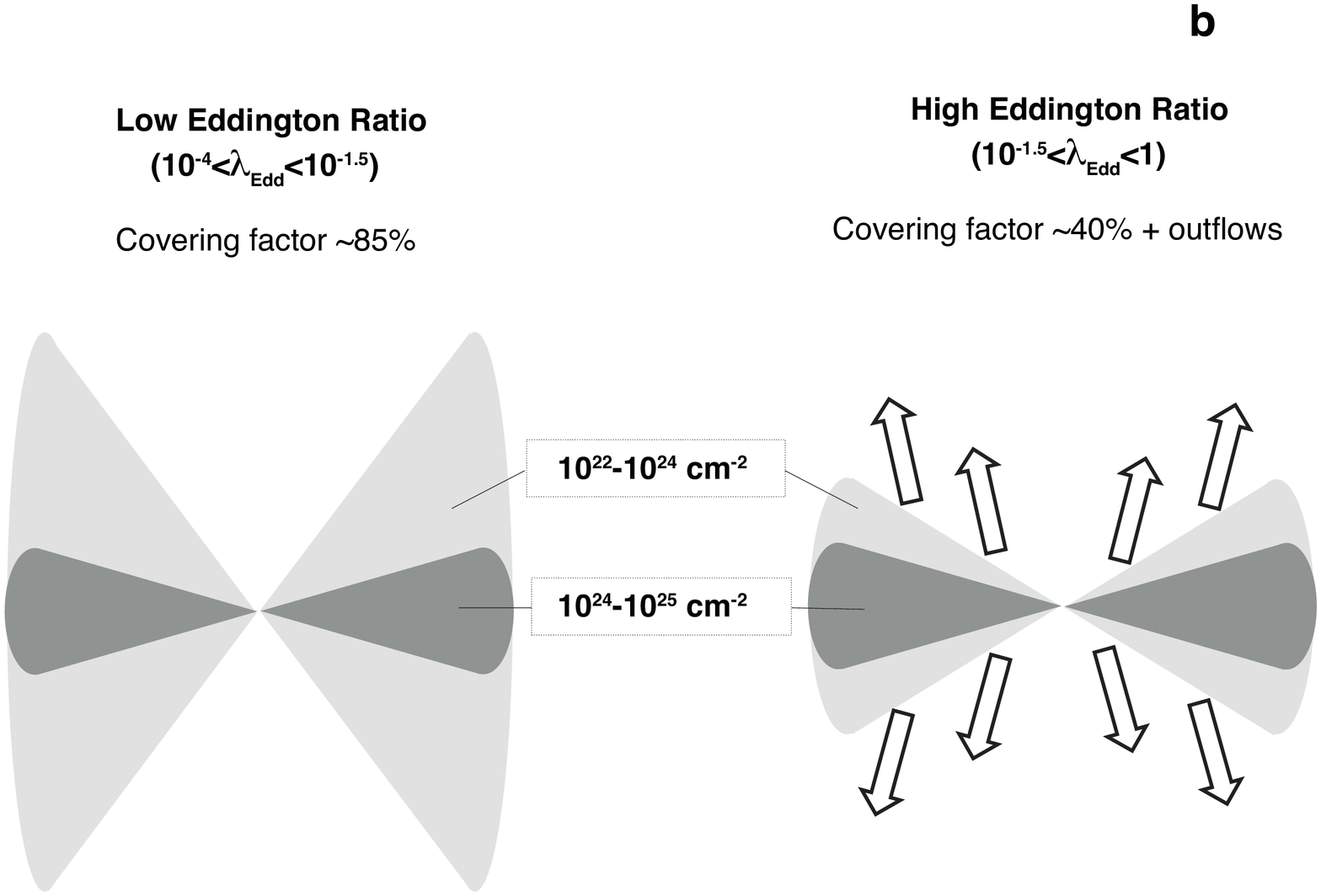}
 \caption{\textbf{Radiation-regulated unification of AGN.} Panel a: relation between the covering factor of dusty gas and the Eddington ratio inferred from our study for both Compton-thick and Compton-thin AGN. In this scheme, defined for $10^{-4} \leq \lambda_{\rm Edd} < 1$, the main parameter regulating the probability of a source being obscured is the Eddington ratio. The black solid curve shows the covering factor of both Compton-thin (yellow area) and Compton-thick (green area) material, while the shaded areas represent the 1$\sigma$ uncertainties.
The covering factor of the material is significantly larger at $\lambda_{\rm Edd}< 10^{-1.5}$ ($\simeq 85\%$, of which $64\pm5\%$ Compton thin and $23\pm 6\%$ Compton thick) than at $\lambda_{\rm Edd}\geq 10^{-1.5}$ ($\simeq 40\%$, of which $19\pm4\%$ Compton thin and $22\pm 7\%$ Compton thick). Panel b: Schematic representation of the material surrounding supermassive black holes for different ranges of Eddington ratio.
 }
 \label{fig:RRUM}
\end{figure}

We calculated the intrinsic column density distribution of AGN in the local Universe by taking into account the selection effects of the {\it Swift}/BAT survey (\S\ref{sec:NHfunction}). We find that, for AGN in the $10^{-4} \leq \lambda_{\rm Edd}< 10^{-1.5}$ range, the fraction of Compton-thick [$N_{\rm H}\geq 10^{24}\rm\,cm^{-2}$] AGN is $f_{\rm CT}=23\pm6\%$, while for those with $10^{-1.5} \leq \lambda_{\rm Edd}< 1$ the value is $f_{\rm CT}=22\pm7\%$. Therefore, while the covering factor of Compton-thin material decreases dramatically with increasing $\lambda_{\rm Edd}$, the Compton-thick material does not follow the same trend. The origin of this difference may be due, at least in part, to the increase of the effective Eddington limit with increasing $N_{\rm H}$. 
In Figure\,\,\ref{fig:RRUM}a we show the evolution of the covering factor with the Eddington ratio, for both Compton-thin and Compton-thick material. This shows that accreting black holes with $10^{-4} \leq  \lambda_{\rm Edd}< 1$ can be generally divided into two types. AGN with $\lambda_{\rm Edd}< 10^{-1.5}$ have obscurers with a large covering factor ($\simeq 85\%$), while those with $\lambda_{\rm Edd}\geq 10^{-1.5}$ have outflowing material and a smaller covering factor ($\simeq 40\%$), half of which is associated with Compton-thick material (see Figure\,\ref{fig:RRUM}b).

Our results imply that the probability of an AGN to be obscured is mostly driven by the Eddington ratio (i.e., a \textit{radiation-regulated unification model}). \noindent This also shows that there is an intrinsic physical difference between obscured and unobscured AGN, i.e. unobscured sources have typically higher $\lambda_{\rm Edd}$. Therefore, studies that adopt obscured AGN as a proxy for the entire AGN population may be inherently biased against highly accreting systems.
The two other parameters regulating the probability of an AGN to be obscured are the inclination angle between the polar axis of the dusty gas structure and the line-of-sight, and whether the host galaxy is in the last stages of a merger or not. Both simulations\cite{Hopkins:2006gf} and recent observations\cite{Satyapal:2014th,Kocevski:2015vh,Ricci:2017bc} have in fact shown that mergers of galaxies can trigger the inflow of material onto the inner parsecs of the system, thus increasing the amount of obscuration.

\noindent The presence of only a small number of obscured sources at $\lambda_{\rm Edd}\geq \lambda_{\rm Edd}^{\rm eff}$ suggests very short timescales for clearing the obscuring material. This, combined with the fact that $f_{\rm obs}$ intrinsically depends on $\lambda_{\rm Edd}$ ($\propto L/M_{\rm BH}$), implies that most of the line-of-sight obscuring material is located within the region in which the gravitational potential of the black hole dominates over that of the host galaxy. If this were not the case then indeed the luminosity would be the main parameter driving the covering factor, and there should be no dependence on $M_{\rm BH}$. However, we cannot exclude that in some cases at least part of the obscuration is related to the host galaxy. 
Considering a typical range of black hole masses ($M_{\rm BH}\simeq 10^6-10^9M_{\odot}$), this suggests that the bulk of obscuring gas and dust is found within $\sim 1.5-65\rm\,pc$ from the black hole (see \S\ref{sect:obsEddratio}), consistent with the compact pc-scale IR-emitting structures found in local AGN by interferometric observations\cite{Jaffe:2004fj}.

\noindent  The results of our analysis may be considered in the framework of an evolutionary scenario for episodic black hole growth. We envisage that an initially inactive black hole (i.e., $\lambda_{\rm Edd}\ll 10^{-4}$) is triggered by an accretion event, and moves to higher $N_{\rm H}$ and $\lambda_{\rm Edd}$, before reaching $\lambda_{\rm Edd}^{\rm eff}$ (white region in Figure\,\,\ref{fig:EddratioNHdiag}) and expelling most of the obscuring material. The AGN would then spend some time as unobscured before consuming all of its remaining accretion reservoir and moving back to a low $\lambda_{\rm Edd}$.
\noindent For $\lambda_{\rm Edd}\gtrsim 0.02-0.06$ most of the Compton-thin dusty gas is expected to be expelled from the vicinity of the black hole in the form of outflows, and could drive the elongated mid-IR emission recently identified by interferometric\cite{Honig:2013wf} and single dish\cite{Asmus:2016uf} observations of AGN. Such outflows could ultimately interact with the host galaxy, potentially affecting star formation and giving rise to the observed relations between supermassive black holes and galaxy bulges\cite{Fabian:2012eq,Kormendy:2013uf}.

\begin{addendum}
 \item This work is dedicated to the memory of our friend and collaborator Neil Gehrels. We are grateful to the referees for their comments, which greatly helped us to improve the quality of the manuscript.
 We acknowledge the work done by the {\it Swift}/BAT team to make this project possible. We thank Makoto Kishimoto, Chin-Shin Chang, Daniel Asmus, Marko Stalevski, Poshak Gandhi and George Privon for valuable discussion. We thank Nathan Secrest for providing us with the stellar masses of the {\it Swift}/BAT sample. This paper is part of the {\it Swift} BAT AGN Spectroscopic Survey.
 This work is sponsored by the Chinese Academy of Sciences (CAS), through a grant to the CAS South America Center for Astronomy (CASSACA) in Santiago, Chile. We acknowledge financial support from FONDECYT 1141218 (CR, FEB), FONDECYT 1160999 (ET), Basal-CATA PFB--06/2007 (CR, ET, FEB), the China-CONICYT fund (CR), the Swiss National Science Foundation (Grant PP00P2\_138979 and PP00P2\_166159, KS), the Swiss National Science Foundation (SNSF) through the Ambizione fellowship grant PZ00P2\textunderscore154799/1 (MK), the NASA ADAP award NNH16CT03C (MK), the National Key R\&D Program of China grant No. 2016YFA0400702 (LH), the National Science Foundation of China grants No. 11473002 and 1721303 (LH), the ERC Advanced Grant Feedback 340442 (ACF), and the Ministry of Economy, Development, and Tourism's Millennium Science Initiative through grant IC120009, awarded to The Millennium Institute of Astrophysics, MAS (FEB). Part of this work was carried out while CR was Fellow of the Japan Society for the Promotion of Science (JSPS) at Kyoto University. This work was partly supported by the Grant-in-Aid for Scientific Research 17K05384 (YU) from the Ministry of Education, Culture, Sports, Science and Technology of Japan (MEXT). We acknowledge the usage of the HyperLeda database (http://leda.univ-lyon1.fr). 
 \item[Author Contributions]  C.R. wrote the manuscript with comments and input from all authors, and performed the analysis. B.T. calculated the bolometric corrections, B.T., M.J.K., K.O. and I.L. analyzed the optical spectra and inferred the black hole masses, C.R. carried out the broad-band X-ray spectral analysis and Y.U. calculated the intrinsic column density distribution of AGN for different ranges of Eddington ratio.
 \item[Author Information] The authors declare that they have no
competing financial interests. Correspondence and requests for materials
should be addressed to C.R. \\
(email: cricci@astro.puc.cl).
\end{addendum}

\newpage

\pagebreak

\captionsetup[figure]{name=Extended Data Figure}
\captionsetup[table]{name=Extended Data Table}

\setcounter{section}{0}
\setcounter{figure}{0}
\setcounter{table}{0}

\noindent \textbf{\huge Methods}

\section{The 70-month {\it Swift}/BAT AGN sample}\label{sec:sample}

The Burst Alert Telescope (BAT\cite{Barthelmy:2005uq,Krimm:2013ys}) on board the {\it Swift} satellite\cite{Gehrels:2004kx} has been carrying out an all-sky survey in the 14--195\,keV energy range since December 2004. The latest release of the {\it Swift}/BAT source catalogue\cite{Baumgartner:2013ee} contains 836 AGN detected in the first 70 months of the mission. Among these, a total of 105 blazars were identified from the latest release of the Rome BZCAT catalogue\cite{Massaro:2015ys} and based on recent literature. In the following we will refer only to the 731 non-blazar AGN, to avoid effects related to beaming and extended X-ray sources. The non-blazar AGN sample is local, with a median redshift and distance of $z\simeq 0.0367$ and $D=161.6$\,Mpc, respectively. The all-sky coverage and hard X-ray selection, together with the fact that the sample is dominated by local AGN, makes it possibly the best existing sample to study the relation between the obscuration and the physical characteristics of these accreting supermassive black holes (SMBHs).
The {\it Swift}/BAT AGN spectroscopic survey (BASS) is a large effort our group has been carrying out to study local hard X-ray selected AGN across the entire electromagnetic spectrum, and in particular in the X-ray and optical regimes (\S\ref{sec:sample_optical}). The first results of BASS include the study of the 55 Compton-thick AGN detected by {\it Swift}/BAT [Ref. \citen{Ricci:2015tg}, see also Ref. \citen{Koss:2016kq,Akylas:2016rw}], the analysis of near-IR emission lines\cite{Lamperti:2017kq}, the study the relation between AGN X-ray emission and high-ionisation optical emission lines [Ref. \citen{Berney:2015uq}, see also Ref. \citen{Ueda:2015mz}], the study of the link between the physical parameters of the accreting SMBH and optical narrow emission lines\cite{Oh:2017db}, and the analysis of the relationship between the X-ray photon index and the Eddington ratio\cite{Trakhtenbrot:2017sf}.

\subsection{X-ray data and spectral analysis.}\label{sec:sample_xray}

X-ray data at softer energies (0.3--10\,keV) are available for 834 out of the 836 sources (99.8\%) reported in the 70-month {\it Swift}/BAT catalogue. A detailed account of the reduction, cross-calibration, and spectral analysis of the entire X-ray dataset is given in Ref. \citen{Ricci:2017aa}, and here we highlight the aspects relevant for the analysis of the non-blazar in the present study.
The X-ray spectral analysis was carried out in the 0.3--150\,keV range by combining the 70-month averaged {\it Swift}/BAT spectra with data below 10\,keV collected using {\it Swift}/XRT {\it XMM-Newton}/EPIC, {\it Chandra}/ACIS, {\it Suzaku}/XIS, and {\it ASCA}/GIS/SIS. 
The 0.3--150\,keV spectra were typically modelled considering the following components: 
i) an absorbed power-law model with a high-energy cutoff, to reproduce the primary X-ray emission (considering both photoelectric absorption and Compton scattering); ii) an unobscured reflection component, which was taken into account by using a slab reflection model (\textsc{pexrav}\cite{Magdziarz:1995pi}); iii) a soft excess component; iv) a Gaussian line to reproduce the Fe K$\alpha$ emission line\cite{Nandra:1994ly,Shu:2010qv,Ricci:2014dq}; v) a cross-calibration constant to account for possible flux variability and calibration uncertainties between the soft X-ray observations and the 70-months averaged hard X-ray emission. For unobscured AGN we used a blackbody model to reproduce the soft excess, while for obscured sources we included a scattered component in the form of a power-law with a cutoff. The normalization and photon index of the scattered emission were fixed to those of the primary X-ray emission, and this component was multiplied by a constant, which was left free to vary and had a typical value of a few percent\cite{Ueda:2007th,Kawamuro:2016fj}. For obscured AGN we also considered, when warranted by the data, emission from collisionally ionized plasma.
We included additional Gaussian features to remove possible residuals in the rest-frame 6--7.5\,keV region and/or below 4\,keV. In order to improve the constraints on the column density and intrinsic luminosity, we fitted the broad-band X-ray spectra of the AGN with $N_{\rm H}$ consistent with $\geq 10^{24}\rm\,cm^{-2}$ within their $90\%$ confidence intervals with a physical torus model\cite{Brightman:2011fe}, which considers absorption and reflection for a spherical-toroidal geometry. To this model we added a scattered component and, if required by the data, additional components of Gaussian lines, a cross-calibration constant and collissionally ionized plasma. Completely unabsorbed AGN [$ N_{\rm H} \lesssim 10^{20}\rm\,cm^{-2}$] were arbitrarily assigned $N_{\rm H} = 10^{20}\rm\,cm^{-2}$. The median of the 90$\%$ confidence interval of the column density is $0.12$\,dex for AGN with $N_{\rm H}=10^{20}-10^{24}\rm\,cm^{-2}$, and $0.35$\,dex for Compton-thick AGN.

The intrinsic luminosity in the 2--10\,keV range (i.e., absorption and k-corrected) was calculated assuming a cosmological model with $H_{0}=70\rm\,km\,s^{-1}\,Mpc^{-1}$, $\Omega_{\mathrm{m}}=0.3$ and $\Omega_{\Lambda}=0.7$. The first results of this work have been reported in Ref.\,\citen{Ricci:2015tg}, where we found that the intrinsic fraction of Compton-thick [$\log (N_{\rm H}/\rm cm^{-2})=24-25$] AGN is $27\pm4\%$, and discussed the intrinsic distribution of $N_{\rm H}$ of AGN in two luminosity bins. The identification of Compton-thick sources in Ref. \citen{Ricci:2015tg} has been recently confirmed by an independent work carried out using a Bayesian approach\cite{Akylas:2016rw}.

\begin{figure}[t!]
 \centering
 \includegraphics[width=0.48\textwidth]{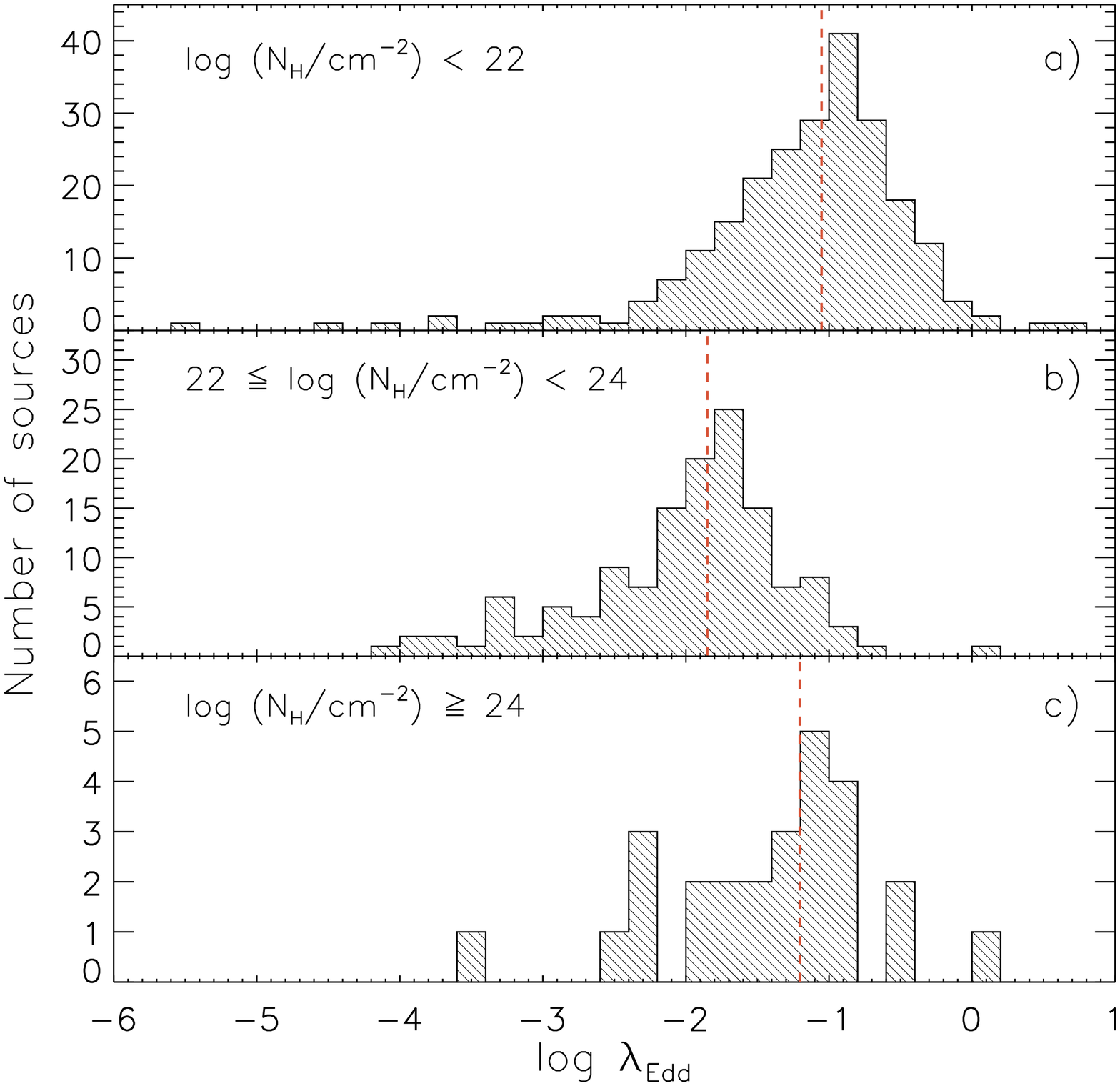}
 \caption{\textbf{Eddington ratio distribution for different classes of AGN.} Histograms of $\lambda_{\rm Edd}$ for unobscured [$N_{\rm H}< 10^{22}\rm\,cm^{-2}$, panel a], obscured Compton-thin [$10^{22} \leq N_{\rm H}< 10^{24}\,\rm cm^{-2}$, panel b] and Compton-thick [$N_{\rm H}\geq 10^{24}\,\rm cm^{-2}$, panel c] AGN. The vertical red dashed lines show the median values for the different subsets of sources.
 }
\label{fig:histogramEddratio}
\end{figure}

\smallskip
\subsection{Optical data and black hole masses.}\label{sec:sample_optical}

Optical spectroscopy has been obtained for 642\,AGN, allowing to infer black hole masses ($M_{\rm BH}$) for 429 non-blazar AGN, of which 232 are unobscured, 166 are Compton-thin obscured and 31 are Compton-thick AGN. We refer the reader to Ref. \citen{Koss:2017aa} for a detailed description of the data and procedures used to derive black hole masses. We briefly outline in the following how we selected the most robust black hole mass estimates for our sources, and the number of objects for which each black hole mass estimation was adopted.

First, for 51 of our AGN, we used literature values reported by studies where $M_{\rm BH}$ was determined through ``direct'' methods, including maser emission (eight AGN), spatially resolved gas- or stellar-kinematics (four and two objects, respectively), and/or reverberation mapping (37 AGN, using the recent compilation of Ref. \citen{Bentz:2015mb}).
Next, for 213 AGN we derived masses from single-epoch spectra of either the broad H$\beta$ (170 objects) or H$\alpha$ (43 sources) emission lines, following prescriptions that are fundamentally based on the results of reverberation mapping. From this subset we removed the 37 sources with $N_{\rm H}\geq 10^{22}\,\rm cm^{-2}$, as their black hole masses are likely underestimated because of extinction of the optical emission, and particularly the continuum and/or the broad Balmer lines.
For H$\beta$, we used the same spectral decomposition method and $M_{\rm BH}$ prescription as described in Ref.\,\citen{Trakhtenbrot:2012hq}. For H$\alpha$, we used the spectral decomposition method described in Ref.\,\citen{Oh:2015if}, and the $M_{\rm BH}$ prescription of Ref.\,\citen{Greene:2005wf}. The uncertainties on these single-epoch $M_{\rm BH}$ determinations are of order $\sim0.3-0.4$ dex [e.g., Ref.\,\citen{Shen:2013ss,Peterson:2014cj} and references therein]. Finally, for 165 objects, $M_{\rm BH}$ was estimated by combining measurements of the stellar velocity dispersion, $\sigma_{*}$, and the $M_{\rm BH}-\sigma_{*}$ relation, using the prescriptions outlined in Ref.\,\citen{Kormendy:2013uf}. For these objects the uncertainties on $M_{\rm BH}$ are also $\sim0.3-0.4$ dex\cite{Gebhardt:2000fj}.
Among the sources with $\log (N_{\rm H}/\rm cm^{-2})<22$ the black hole masses were obtained using broad H$\beta$ (156), reverberation mapping (35), broad H$\alpha$ (20), velocity dispersion (19), stellar (1) and gas (1) kinematics. For the objects with $\log (N_{\rm H}/\rm cm^{-2})\geq 22$ we used velocity dispersion (146), masers (8), reverberation mapping (2), and gas (3) and stellar (1) kinematics.
The objects with black hole masses are a representative subsample of all {\it Swift}/BAT detected AGN: performing a Kolmogorov-Smirnov test on the two distributions of the intrinsic 14-150\,keV luminosity we obtain a p-value of $\simeq 0.98$.

%*******************************************************************************************************************************************************************************

\begin{figure}[t!]
\centering
\includegraphics[width=0.45\textwidth]{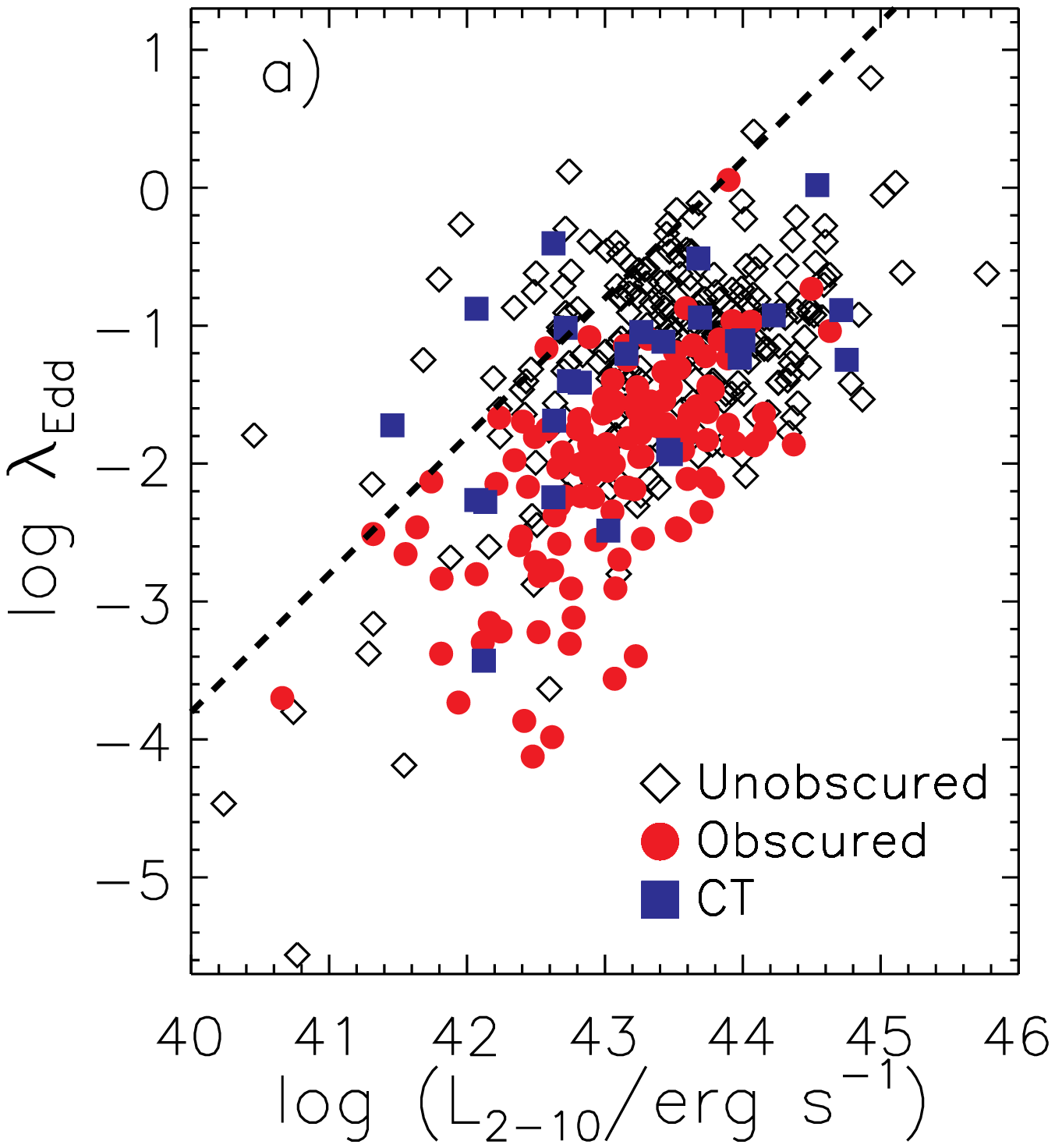}
\par\bigskip
\includegraphics[width=0.45\textwidth]{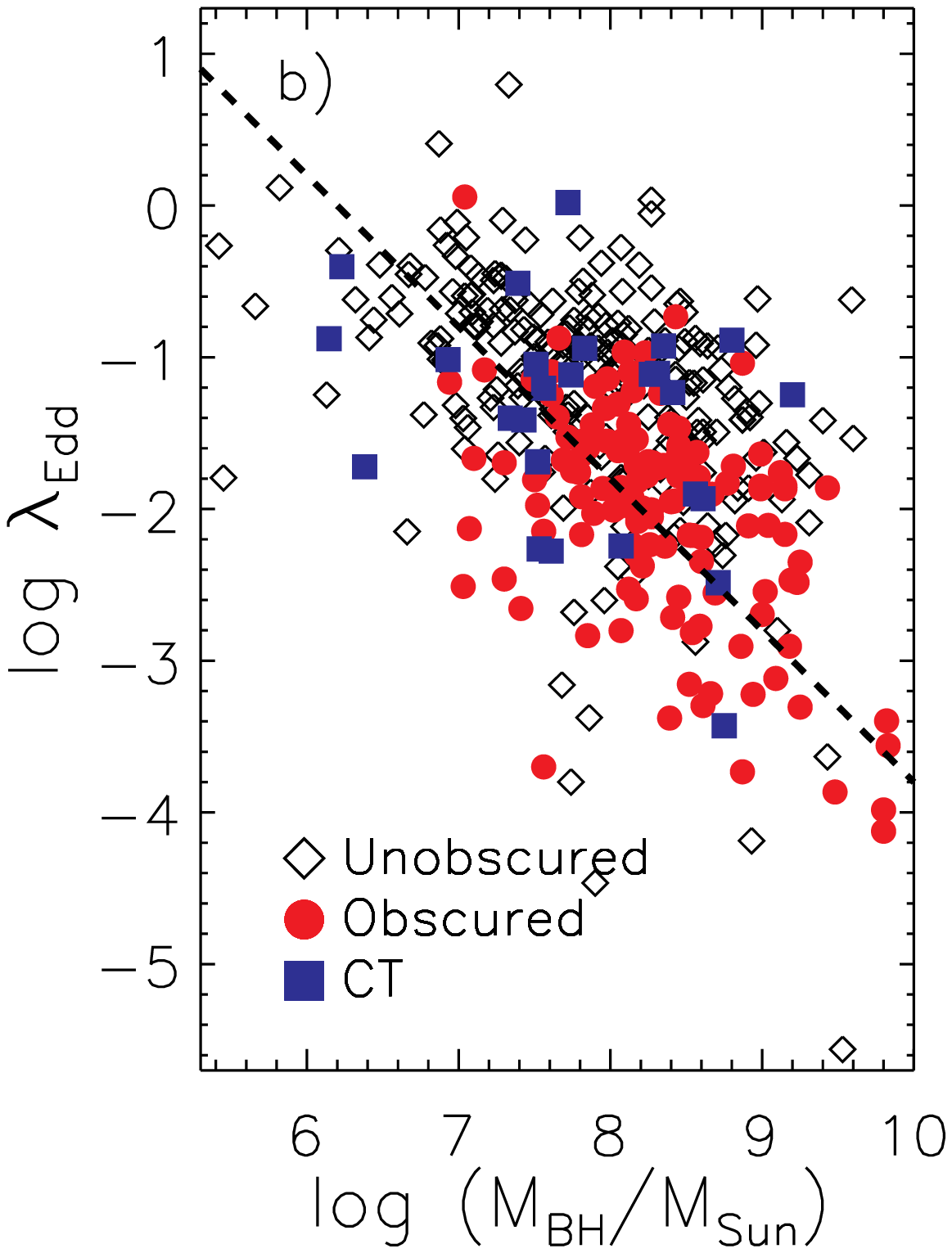}
\caption{\textbf{Eddington ratio versus X-ray luminosity and black hole mass.} Scatter plots of $\lambda_{\rm Edd}$ versus the 2--10\,keV intrinsic luminosity (panel a) and the black hole mass (panel b) for unobscured [$N_{\rm H} \leq 10^{22}\rm\,cm^{-2}$, black empty diamonds], obscured [$10^{22} \leq N_{\rm H}< 10^{24}\rm\,cm^{-2}$, red filled circles] and Compton-thick [$N_{\rm H}\geq 10^{24}\rm\,cm^{-2}$, blue filled squares] AGN. The black dashed lines represent values for constant mass (panel a) and luminosity (panel b). }
\label{fig:EddratiovsL_MBH}
\end{figure}

\section{The Eddington limit for dusty gas and the $N_{\rm H}-\lambda_{\rm Edd}$ diagram}\label{sec:EddNHdiagram}

The Eddington limit is defined as the luminosity at which the radiation pressure from a source, in this case the accreting SMBH, balances the gravitational attraction. This is typically determined using the following equation:
\begin{equation}\label{eq:eddratio}
L_{\rm Edd}=\frac{4\pi G M m_{\rm p}c}{\sigma_{\rm T}},
\end{equation}
where $G$ is the gravitational constant, $M$ is the mass of the system (in this case $M=M_{\rm BH}$), $m_{\rm p}$ is the mass of the proton, $c$ is the speed of light, and $\sigma_{\rm T}$ is the Thomson cross-section. The Eddington ratio is the ratio between the bolometric and the Eddington luminosity ($\lambda_{\rm Edd}=L_{\rm Bol}/L_{\rm Edd}$). Here we calculated the bolometric luminosity from the intrinsic (i.e. absorption and k-corrected) 2--10\,keV luminosity, using a 2--10\,keV bolometric correction $\kappa_{2-10}=20$ (Ref.\,\citen{Vasudevan:2009ng}). We conservatively estimate that the typical uncertainty on $\lambda_{\rm Edd}$ is $\sim 0.5$\,dex. The Eddington ratio distribution of the sources of our sample divided into unobscured [$N_{\rm H} \leq 10^{22}\rm\,cm^{-2}$], obscured [$10^{22} \leq N_{\rm H}< 10^{24}\rm\,cm^{-2}$] and Compton-thick [$N_{\rm H}\geq 10^{24}\rm\,cm^{-2}$] AGN is shown in Extended Data Figure\,\,\ref{fig:histogramEddratio}, while in Extended Data Figure\,\,\ref{fig:EddratiovsL_MBH} we show the scatter plot of $\lambda_{\rm Edd}$ versus the 2-10\,keV luminosity ($L_{2-10}$; panel a) and $M_{\rm BH}$ (panel b).

The cross-section of the process used in Equation\,\ref{eq:eddratio} is that of the photon-electron scattering (i.e. the Thomson cross-section), so that the Eddington limit is defined for ionized hydrogen. However, if the material includes both dust and gas, then the effective cross-section\cite{Fabian:2006lq} of the interaction between radiation and matter, $\sigma_{\rm i}$, is larger, $\sigma_{\rm i}> \sigma_{\rm T}$. The effective cross section is averaged over the incident spectrum, and depends on the physical properties of the material (i.e., $N_{\rm H}$, ionization state, chemical composition, dust content). 
Therefore, for dusty gas, the effective Eddington limit would be lower\cite{Fabian:2006lq,Honig:2007sh} than that obtained using $L_{\rm Edd}$ from Equation\,\ref{eq:eddratio}. One can define the effective Eddington limit as $\lambda_{\rm Edd}^{\rm eff}=1/A$, where the boost factor\cite{Fabian:2006lq} $A$ can simply be thought as the ratio between the effective, frequency-weighted absorption cross-section and the cross-section for ionized hydrogen: $A=\sigma_{\rm i}/\sigma_{\rm T}$. For $\lambda_{\rm Edd}\geq \lambda_{\rm Edd}^{\rm eff}$ the force exerted by the radiation field surpasses that of the gravitational pull.

Using \textsc{cloudy}\cite{Ferland:1993os} it has been shown\cite{Fabian:2006lq,Fabian:2008hc,Fabian:2009ez} that, for dusty gas, the boost factor ranges from several hundred (for $N_{\rm H}\sim 10^{20}\rm\,cm^{-2}$) to unity (for $N_{\rm H}\simeq 10^{24}\rm\,cm^{-2}$). 
Since dust absorption is greater in the UV, this portion of the SED has the strongest influence in driving the material away. However, in the UV the material becomes optically thick already at $N_{\rm H}\simeq 10^{21}\rm\,cm^{-2}$, so that the outer layers are pushed by the inner ones. As explained by Ref.\,\citen{Fabian:2008hc}, the X-ray emission keeps the gas and dust weakly ionized, ensuring that they are effectively bound by Coulomb forces and that the pressure on the dust grains is also exerted on the (partially) ionized gas.

It should be noted that the accretion disc is thought to emit anisotropically, with the frequency-specific flux density decreasing with higher inclination angles ($F\propto \cos \theta_{\rm i}$). Moreover, the UV radiation, which constitutes most of the bolometric output of the AGN, may be even more strongly collimated than the optical radiation (i.e., due to limb darkening). Thus, one would expect that also the radiation pressure exerted on the dusty gas has a similar dependency on the inclination angle\cite{Liu:2011kq}. This could contribute to the anisotropy of the obscuring material, since radiation pressure would preferentially remove gas and dust located pole-on with respect to the accretion disk.
The values of $\lambda_{\rm Edd}^{\rm eff}$ for different grain compositions have been calculated by Ref.\,\citen{Fabian:2009ez}, who showed that, considering a grain abundance typical of the interstellar medium, the accreting SMBH would be able to push away material with a column density of $N_{\rm H}\simeq 10^{22}\rm\,cm^{-2}$ ($N_{\rm H}\simeq 10^{23}\rm\,cm^{-2}$) already at $\lambda_{\rm Edd}\simeq 0.02$ ($\lambda_{\rm Edd}\simeq 0.15$). The increase of $\lambda_{\rm Edd}^{\rm eff}$ with $N_{\rm H}$ is due to the larger weight radiation pressure needs to push away.

\begin{figure}[t!]
 \centering
\includegraphics[width=0.48\textwidth]{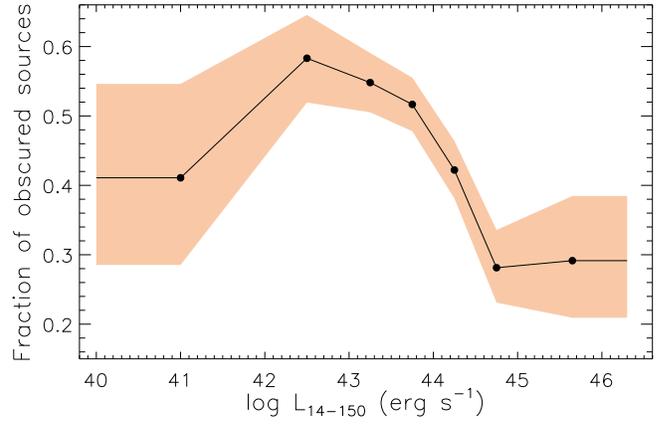}
 \caption{\textbf{Fraction of obscured sources versus luminosity.} Fraction of obscured Compton-thin sources versus the intrinsic 14--150\,keV luminosities for the non-blazar AGN of the {\it Swift}/BAT 70-month catalog. The fraction of obscured sources is normalized in the $N_{\rm H}=10^{20}-10^{24}\rm\,cm^{-2}$ range. The filled area represents the 16th and 84th quantiles of a binomial distribution\cite{Cameron:2011cl}.
 }
 \label{fig:fobs_vs_luminosity}
\end{figure}

One can define a {\it forbidden} (or {\it blowout}) region in the $N_{\rm H}-\lambda_{\rm Edd}$ diagram where long-lived clouds of dusty gas cannot exist, due to radiation pressure\cite{Fabian:2008hc,Fabian:2009ez,Raimundo:2010cl,Fabian:2012eq,Vasudevan:2013fu}. The $N_{\rm H}-\lambda_{\rm Edd}$ diagram for the sources of our sample is shown in Figure\,\,\ref{fig:EddratioNHdiag}, together with the theoretical values of $\lambda_{\rm Edd}^{\rm eff}$ for dusty gas\cite{Fabian:2009ez} as a function of $N_{\rm H}$. We extend here the effective Eddington ratio to $\log \lambda_{\rm Edd}\geq 0$ by linearly extrapolating $\lambda_{\rm Edd}^{\rm eff}(N_{\rm H})$ from $N_{\rm H} \leq 10^{24}\rm\,cm^{-2}$ (Ref.\,\citen{Fabian:2009ez}) to $N_{\rm H} > 10^{24}\rm\,cm^{-2}$. We find that most of the sources of our sample are well within the limits expected by radiation pressure on dusty gas, and tend to lie outside the forbidden region (the white area in the figure).

\begin{figure*}[t!]
 \centering
\includegraphics[width=0.48\textwidth]{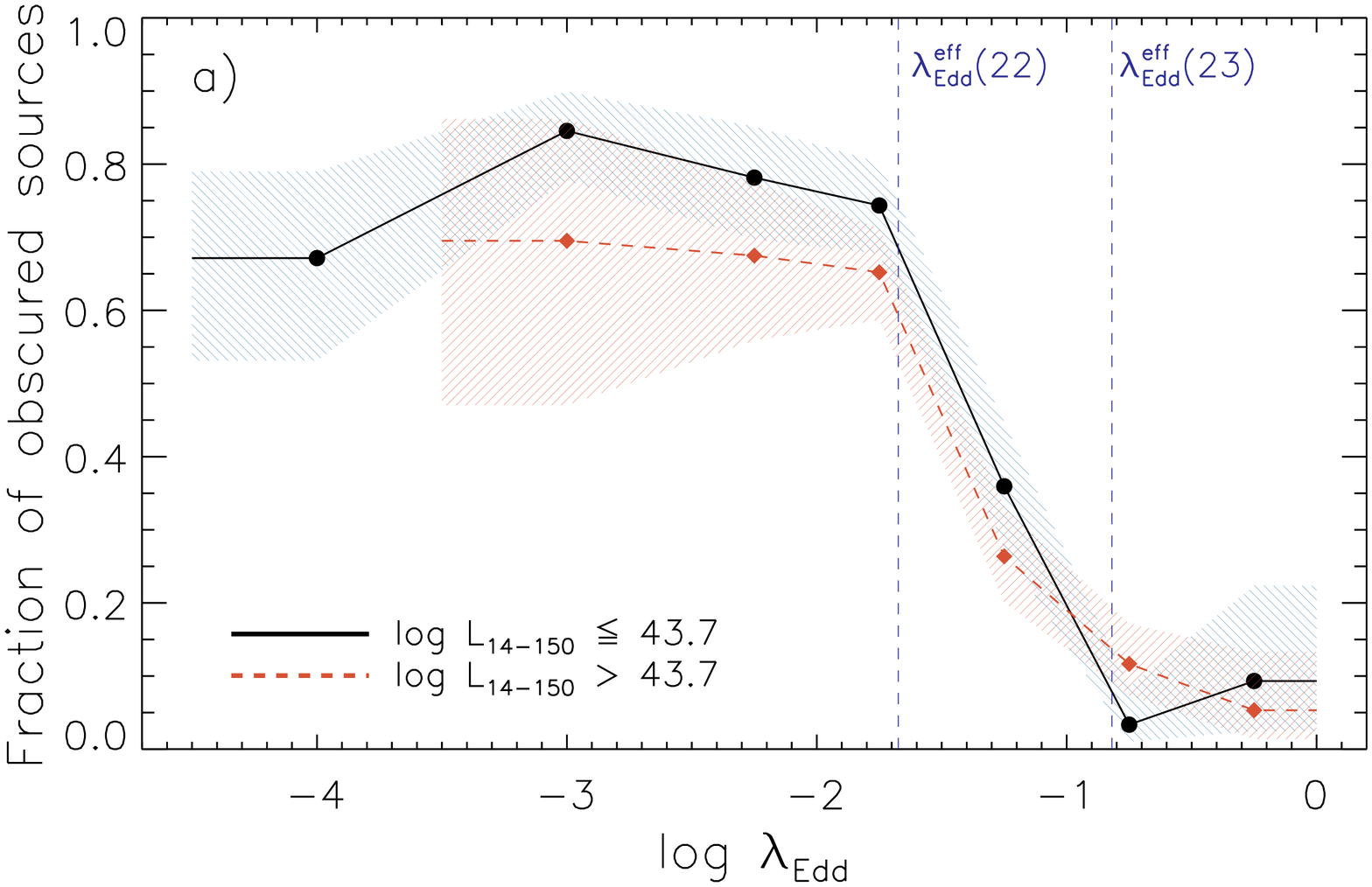}
%\par\bigskip
\includegraphics[width=0.48\textwidth]{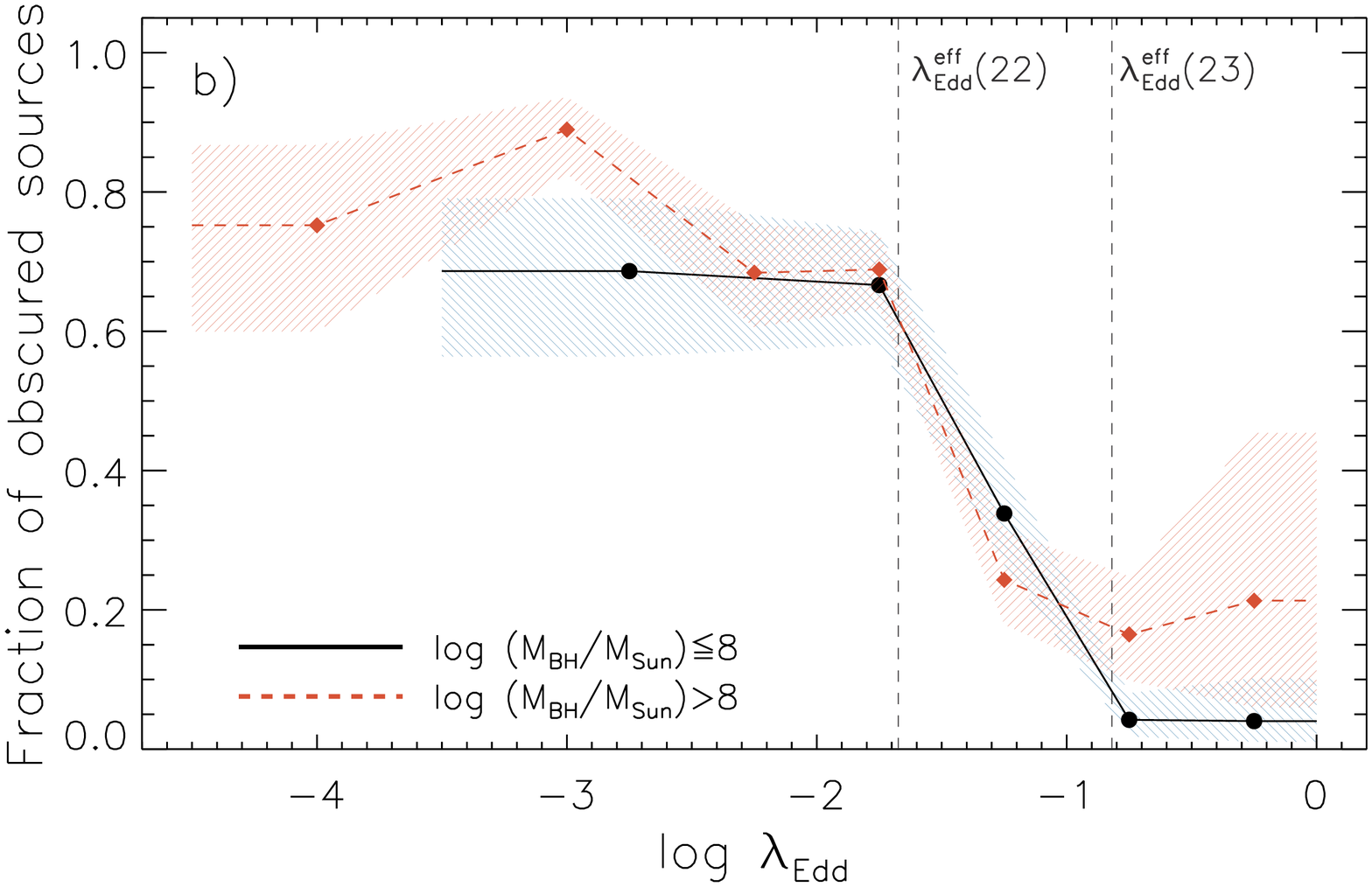}
 \caption{\textbf{Fraction of obscured sources versus $\lambda_{\rm Edd}$ for two ranges of luminosity and black hole mass.} Fraction of obscured Compton-thin sources versus Eddington ratio for two bins of the 14--150\,keV intrinsic luminosity (panel a) and of black hole mass (panel b). The dashed vertical lines represent the effective Eddington limit for dusty gas with $N_{\rm H}=10^{22}\rm\,cm^{-2}$ [$\lambda_{\rm Edd}^{\rm eff}(22)$] and $N_{\rm H}=10^{23}\rm\,cm^{-2}$ [$\lambda_{\rm Edd}^{\rm eff}(23)$] for standard dust grain composition of the interstellar medium. The plots are normalized to unity in the $20 \leq \log (N_{\rm H}/\rm cm^{-2})< 24$ interval, and the shaded areas represent the 16th and 84th quantiles of a binomial distribution\cite{Cameron:2011cl}. The figures clearly show that the same trend found for the whole sample is obtained when looking at different bins of $L_{14-150}$ and $M_{\rm BH}$, confirming that the Eddington ratio is the main parameter driving obscuration.
 }
 \label{fig:FobsEddratio_Lbins}
\end{figure*}

%*******************************************************************************************************************************************************************************

\section{The covering factor of the Compton-thin obscuring material and its relation with luminosity and Eddington ratio}\label{sec:obsvsEddratio}

\subsection{Obscuration and luminosity.}\label{sect:obsLuminosity}

The fraction of obscured Compton-thin sources ($f_{\rm obs}$) is directly connected to the covering factor of the obscuring material surrounding the accreting SMBH. The first evidence of a decrease of $f_{\rm obs}$ with AGN luminosity was obtained studying the ratio between type-I and type-II AGN (i.e., objects with and without broad optical emission lines, respectively) more than 30 years ago\cite{Lawrence:1982ys}. 
This trend was confirmed by successive optical studies\cite{Lawrence:1991vn,Simpson:2005uu}, while X-ray studies have clearly shown that the fraction of Compton-thin obscured sources decreases with the luminosity\cite{Ueda:2003qf,La-Franca:2005uf,Sazonov:2007if,Hasinger:2008ve,Della-Ceca:2008xu,Beckmann:2009fk,Ueda:2011fk,Brightman:2011rw,Ueda:2014ix,Merloni:2014qv,Buchner:2015ve,Aird:2015gf,Georgakakis:2017jk} with a slope of $f_{\rm obs} \propto -0.226 \log (L_{2-10}/\rm\,erg\,s^{-1})$ in the $42 \leq \log (L_{2-10}/\rm\,erg\,s^{-1}) \leq 46$ range\cite{Hasinger:2008ve}. This trend was also shown to reproduce the observed decrease of the Fe\,K$\alpha$ equivalent width with increasing luminosity\cite{Ricci:2013cz} (i.e., the X-ray Baldwin effect\cite{Iwasawa:1993yb,Bianchi:2007et,Ricci:2014dg}), and it has been shown to be due to the intrinsically different space densities of obscured and unobscured AGN for a given luminosity\cite{Tueller:2008yg,Burlon:2011dk}.
In the IR, the covering factor of the dust can be inferred as the ratio between the mid-IR and the bolometric luminosities. This is driven by the idea that the fraction of the bolometric AGN luminosity reprocessed by the dust surrounding the accreting system is proportional to the covering factor of the dusty absorber. Similarly to what was found in the optical and X-ray bands, several IR studies have shown that the covering factor of the dust decreases with the bolometric luminosity\cite{Maiolino:2007ii,Treister:2008ff,Gandhi:2009uq,Assef:2013bu,Lusso:2013fy,Toba:2014ao,Lacy:2015rm,Stalevski:2016kl,Mateos:2016ys,Ichikawa:2017xr}. We note, however, that several other IR studies of high-luminosity AGN (including at high redshifts) do {\it not} identify such a trend\cite{Netzer:2016aa}.

The covering factor of the obscuring material in the IR can also be obtained by fitting the SED with torus models\cite{Honig:2006tg,Nenkova:2008kl,Nenkova:2008lg,Schartmann:2008kv,Honig:2010tt,Honig:2010vp,Stalevski:2012kq,Siebenmorgen:2015br}, and analyses performed using clumpy torus models have confirmed the existence of a decrease of the covering factor of the dust for increasing luminosities\cite{Mor:2009fk,Alonso-Herrero:2011jx}, while showing that typically the tori of type-II AGN have a larger number of clouds than those of type-I AGN\cite{Ramos-Almeida:2011eb}. This is likely to be a selection effect: considering a randomly drawn inclination angle, sources with large tori and high numbers of clouds are more likely to be observed as type-II AGN\cite{Elitzur:2012dq}.

\begin{figure}[h!]
 \centering
\includegraphics[width=0.48\textwidth]{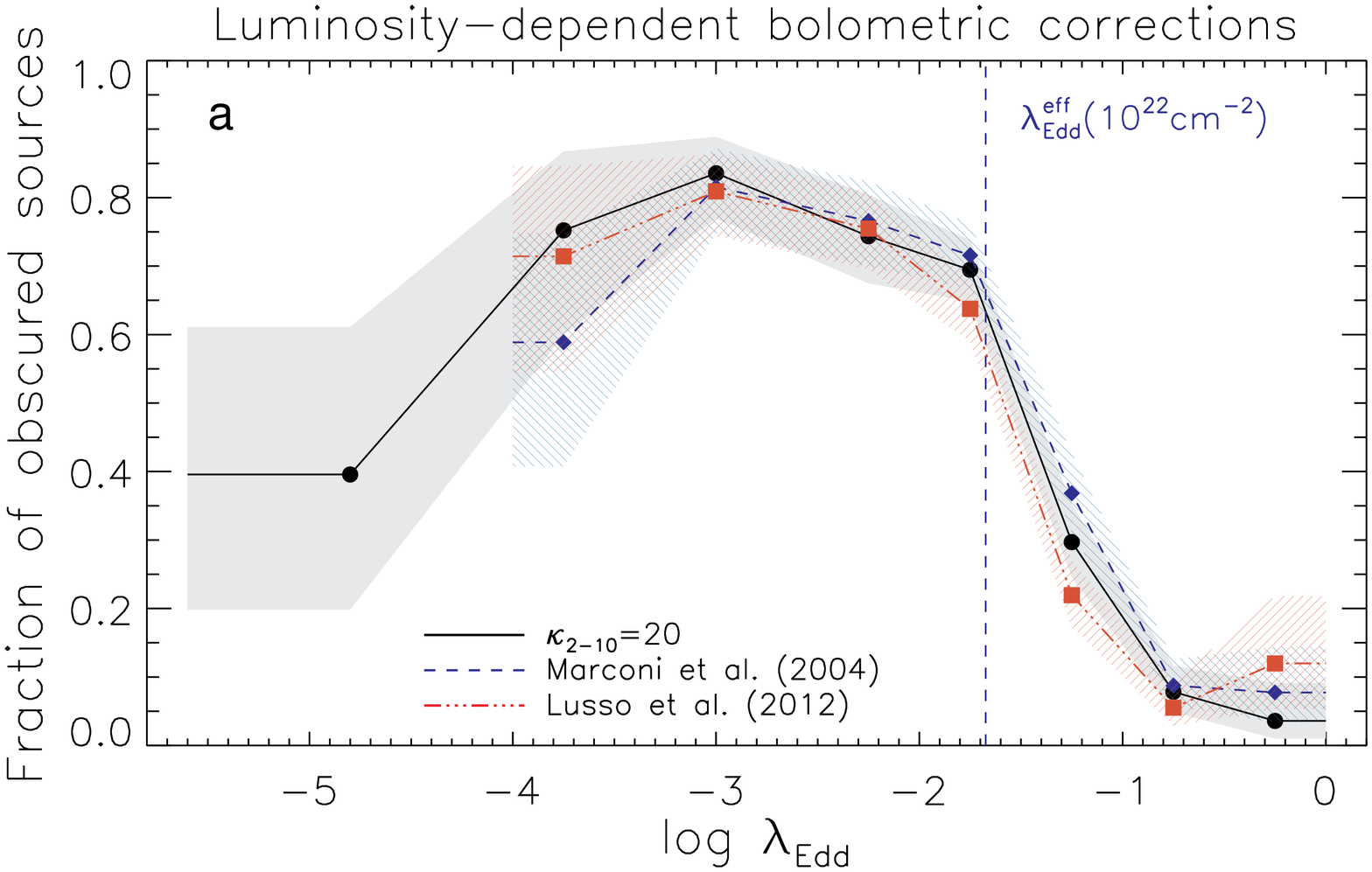}
\par\bigskip
\includegraphics[width=0.48\textwidth]{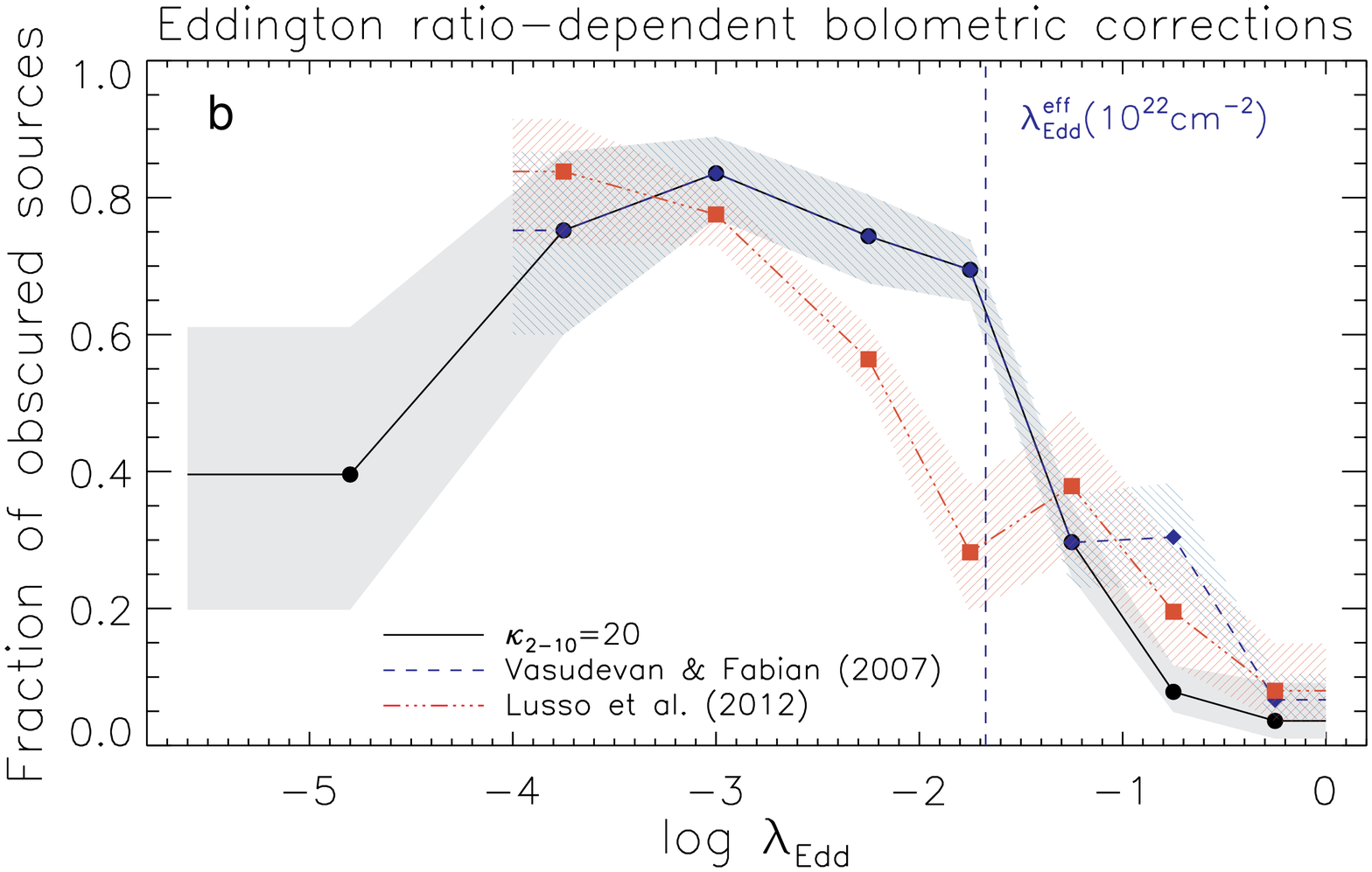}
 \caption{\textbf{Relation between the fraction of obscured AGN and the Eddington ratio assuming different bolometric corrections}. The bolometric corrections used are dependent on the bolometric luminosity (panel a) and on the Eddington ratio (panel b). The shaded areas represent the 16th and 84th quantiles of a binomial distribution\cite{Cameron:2011cl}. The figure shows that our results are mostly independent on the choice of the bolometric correction.   
 }
 \label{fig:fobs_vs_Eddratio_diffBolcorrections}
\end{figure}

\begin{figure}[h!]
 \centering
\includegraphics[width=0.48\textwidth]{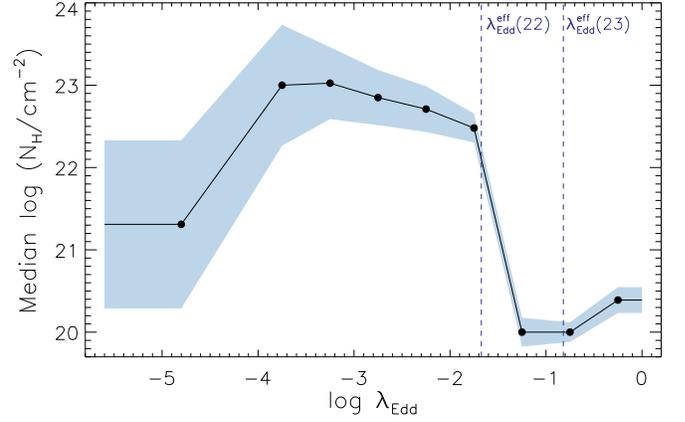}
 \caption{\textbf{Median value of the column density versus Eddington ratio for AGN with $20\leq \log (N_{\rm H}/\rm cm^{-2}) \leq 24$.} The plot highlights the sharp transition at $\log \lambda_{\rm Edd}\simeq -1.5$ between AGN being typically significantly obscured to unobscured. The filled area shows the median absolute deviation. The dashed vertical lines represent the effective Eddington limit for a dusty gas with $N_{\rm H}=10^{22}\rm\,cm^{-2}$ [$\lambda_{\rm Edd}^{\rm eff}(22)$] and $N_{\rm H}=10^{23}\rm\,cm^{-2}$ [$\lambda_{\rm Edd}^{\rm eff}(23)$] for standard dust grain composition of the interstellar medium, showing that radiation pressure regulates the median column density of AGN.
 }
 \label{fig:medianNH_vs_Eddratio}
\end{figure}

The receding torus model\cite{Lawrence:1991vn}, which was proposed to explain the decrease of the covering factor of the absorber with the luminosity, foresees that the inner radius of the obscuring material is set by the dust sublimation radius\cite{Suganuma:2006ct,Kishimoto:2007jb,Kishimoto:2011ax}, which scales with luminosity following $R_{s}\propto L^{0.5}$. Assuming that the height of the torus does not change, this model predicts that the fraction of the accreting system covered by the obscuring material would be $f_{\rm obs}\propto L^{-0.5}$. However, X-ray studies have shown that the slope of the $f_{\rm obs}-L$ trend is flatter than expected by this model\cite{Burlon:2011dk}. Moreover, using optically-selected sources it has been recently argued that the relation between the fraction of type-I AGN and the luminosity appears to be more complex, and that the Eddington ratio might play a role\cite{Oh:2015if}. X-rays are absorbed by both dust and gas, so that pushing further away the sublimation radius for increasing luminosities may not be enough to affect X-ray obscuration. It has been recently proposed that, while in low-luminosity sources most of the absorption could be related to the dusty torus, in AGN with high luminosity part of the obscuration might be due to dust-free clouds in the broad-line region located well within the sublimation region\cite{Davies:2015ve}. These clouds would obscure only the X-ray emission and not the optical radiation, and could explain the existence of sources that are optically unobscured but show evidence of absorption in their X-ray spectra\cite{Merloni:2014qv}.
% %
%
%
In a recent work carried out using a physical IR torus model\cite{Stalevski:2016kl}, it has been shown that the anisotropic emission of the torus should be taken into account when computing the dust covering factors, and would flatten the decline with luminosity. Similarly, it has been proposed\cite{Sazonov:2015dq} that the anisotropic reprocessed X-ray radiation from the torus might affect the relation between $f_{\rm obs}$ and the X-ray luminosity. 
Several of the outstanding issues with the X-ray-through-IR unification scheme of AGN, and particularly the receding torus model, are discussed in recent reviews\cite{Netzer:2015rev,Brandt:2015kq}.

From our complete sample of {\it Swift}/BAT non-blazar AGN we find a behaviour of $f_{\rm obs}$ similar to what has been found by previous studies of local AGN (Extended Data Figure\,\,\ref{fig:fobs_vs_luminosity}; see also Refs.\,\citen{Beckmann:2009fk,Burlon:2011dk,Kawamuro:2016qv}), with a peak at $\log (L_{14-150}/\rm\,erg\,s^{-1})\simeq 43$, a rather steep decline up to $\log (L_{14-150}/\rm\,erg\,s^{-1})\simeq 45$, and a tentative decrease at $\log (L_{14-150}/\rm\,erg\,s^{-1})\lesssim 42$.

\subsection{Obscuration and Eddington ratio.}\label{sect:obsEddratio}

While the relation between the covering factor and the luminosity has been the subject of intense study over the past decades, very little is currently known about the effect of the Eddington ratio on obscuration. Possible evidence of a difference in the $\lambda_{\rm Edd}$ distributions of obscured and unobscured AGN has been found by earlier studies carried out using {\it Swift}/BAT\cite{Winter:2009xi} and {\it INTEGRAL} IBIS/ISGRI\cite{Beckmann:2009fk}, and from the analysis of AGN at higher redshifts in the Cosmic Evolution Survey field\cite{Lusso:2012it}. More recently, it has been suggested\cite{Buchner:2017gf} that the $f_{\rm obs}-\lambda_{\rm Edd}$ relation might have a shape similar to that of the $f_{\rm obs}-L_{2-10}$ relation, and that the Eddington ratio might drive the covering factor of dust\cite{Ezhikode:2016ty}. However, none of the studies carried out so far was able to disentangle the lower average $\lambda_{\rm Edd}$ of obscured AGN from the underlying differences in the luminosity functions of obscured and unobscured AGN.

The dependence of the $f_{\rm obs}$ on the luminosity disappears when dividing the sample into different bins of $\lambda_{\rm Edd}$. This is confirmed by performing a linear fit on $f_{\rm obs}(L)$, which results in slopes of $-0.05\pm0.43$ and $0.02\pm0.06$, for the $\lambda_{\rm Edd}< 10^{-1.5}$ and $10^{-1.5} \leq \lambda_{\rm Edd}< 1$ bin, respectively. The disappearance of a strong correlation between $f_{\rm obs}$ and the luminosity is also found using different thresholds of $\lambda_{\rm Edd}$ (e.g., $10^{-1.3}$ and $10^{-1.7}$).
Performing a linear fit on $f_{\rm obs}(\lambda_{\rm Edd})$ in the $-3 \leq \log \lambda_{\rm Edd}<0$ range we obtained a p-value of 0.003, showing the existence of a significant correlation.
The strong dependence of the fraction of obscured sources on $\lambda_{\rm Edd}$ is confirmed by the fact that dividing our sample into different ranges of the intrinsic 14--150\,keV luminosity or black hole mass (panels a and b of Extended Data Figure\,\,\ref{fig:FobsEddratio_Lbins}, respectively) all subsets of sources show the same trend observed for the whole sample. 
The same result is obtained when using different thresholds of luminosity (e.g., $10^{44}\rm\,erg\,s^{-1}$ and $10^{44.2}\rm\,erg\,s^{-1}$), black hole mass [e.g., $\log (M_{\rm BH}/M_{\odot})> 8.5$ and $\log (M_{\rm BH}/M_{\odot})\leq 8$], or removing objects with luminosities $<10^{42.5}\rm\,erg\,s^{-1}$.
The fraction of obscured sources decreases more steeply in the range $10^{-1.75} \leq \lambda_{\rm Edd} \leq 10^{-0.75}$  than in the range $10^{-3} \leq \lambda_{\rm Edd} \leq 10^{-1.75}$.
Performing a linear fit we find $f_{\rm obs}=(-0.390\pm0.068)-(0.606\pm0.056)\times\log \lambda_{\rm Edd}$ and $f_{\rm obs}=(0.496\pm0.136)-(0.113\pm0.060)\times\log \lambda_{\rm Edd}$ for the former and latter intervals in $\lambda_{\rm Edd}$, respectively. The similar linear fit over the range $10^{-5.6} \leq \lambda_{\rm Edd}\leq 10^{-2.5}$  yields $f_{\rm obs}=(1.48\pm0.35)+(0.21\pm0.11)\times \log \lambda_{\rm Edd}$.  In calculating the fractions of obscured Compton-thin AGN we took into account the fact that we have a slightly larger number of unobscured AGN than obscured AGN for which the black hole mass was estimated.
We find a significant positive correlation (p-value=0.0064, slope $0.20\pm0.07$) between $N_{\rm H}$ and $\lambda_{\rm Edd}$ for objects with $N_{\rm H}\geq 10^{22}\rm\,cm^{-2}$. This is likely, at least in part, a selection effect, consistent with the idea that larger column densities are necessary to withstand increasing radiation pressure.

We verified whether the presence of mergers might affect our results, since the AGN in these objects might be significantly more obscured than those in isolated galaxies\cite{Satyapal:2014th,Kocevski:2015vh,Koss:2016kq,Ricci:2017bc}. In order to be conservative, we removed from our sample the 32 AGN found in interacting systems (e.g., Ref.\,\citen{Koss:2010qv}), including galaxies both in early and late merger stages. We found for this subsample a relation between obscuration and Eddington ratio consistent with what we obtained for the whole sample. Similarly, the relation between $f_{\rm obs}$ and the luminosity disappears for this subsample when dividing into different bins of $\lambda_{\rm Edd}$.

Our results are largely unaffected by the bolometric corrections adopted. Considering luminosity-dependent bolometric corrections\cite{Marconi:2003cz,Lusso:2012it} we find the same sharp decrease in the fraction of obscured sources at $\lambda_{\rm Edd}\simeq \lambda_{\rm Edd}^{\rm\,eff}(10^{22}\rm\,cm^{-2})$ (Extended Data Figure\,\,\ref{fig:fobs_vs_Eddratio_diffBolcorrections}a). Similarly, using the Eddington-ratio dependent bolometric corrections of Ref.\,\citen{Vasudevan:2007qt} we find results in very good agreement with what was obtained assuming $\kappa_{2-10}=20$ (Extended Data Figure\,\,\ref{fig:fobs_vs_Eddratio_diffBolcorrections}b). Assuming the $\lambda_{\rm Edd}$-dependent bolometric corrections of Ref.\,\citen{Lusso:2012it} we find that the decrease of $f_{\rm obs}$ is found at lower Eddington ratios ($\lambda_{\rm Edd}\simeq 10^{-3}$). This is due to the fact that the bolometric correction at low $\lambda_{\rm Edd}$ is extremely small ($\kappa_{2-10}\simeq4$). Moreover, using these corrections we find that 70 sources are accreting at $\lambda_{\rm Edd}>1$, which corresponds to $\sim 18\%$ of our sample. Such a large fraction of super Eddington sources in our local AGN sample is rather unlikely.

\begin{figure*}[t!]
 \centering
\includegraphics[width=0.48\textwidth]{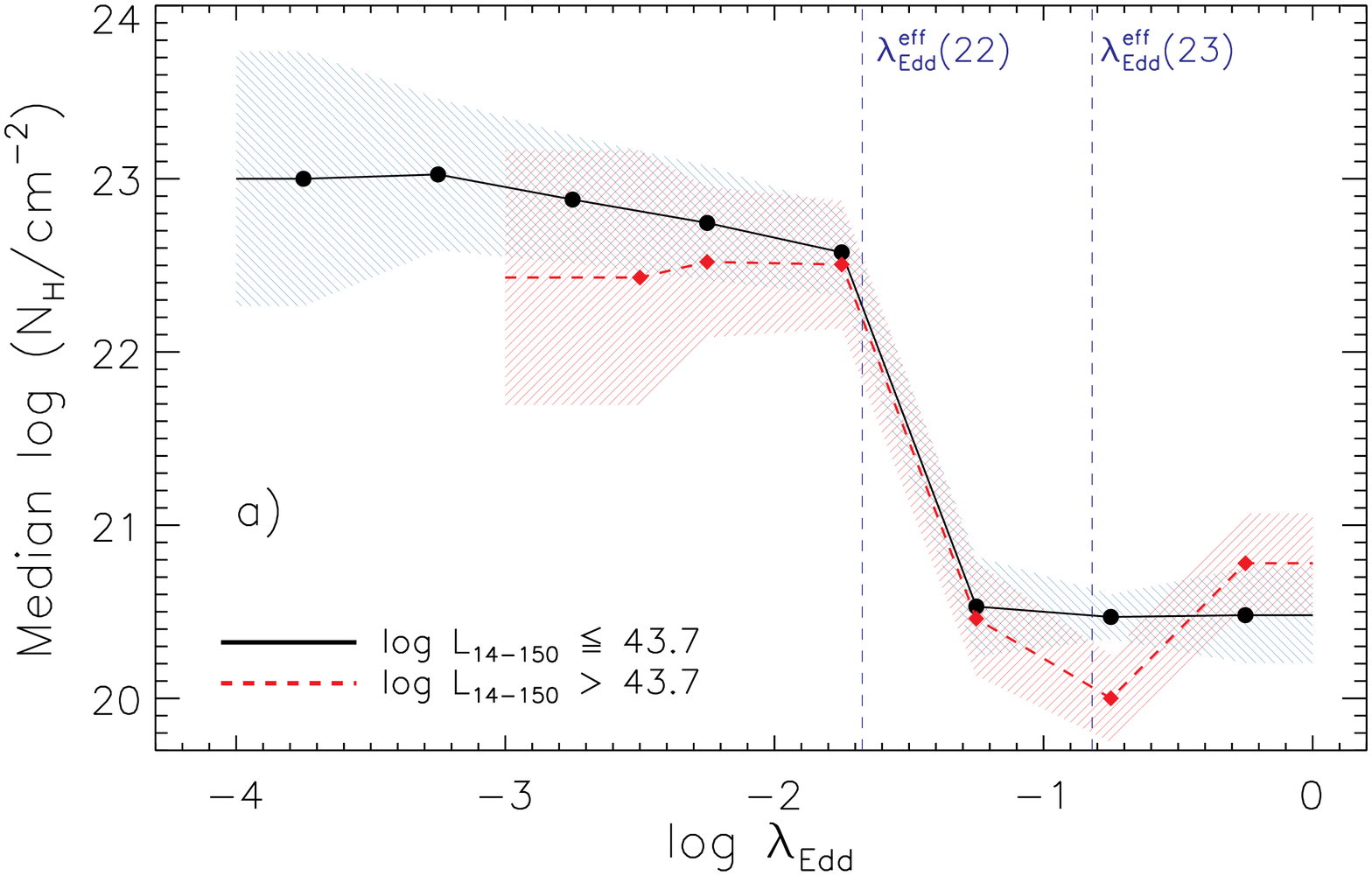}
%\par\bigskip
\includegraphics[width=0.48\textwidth]{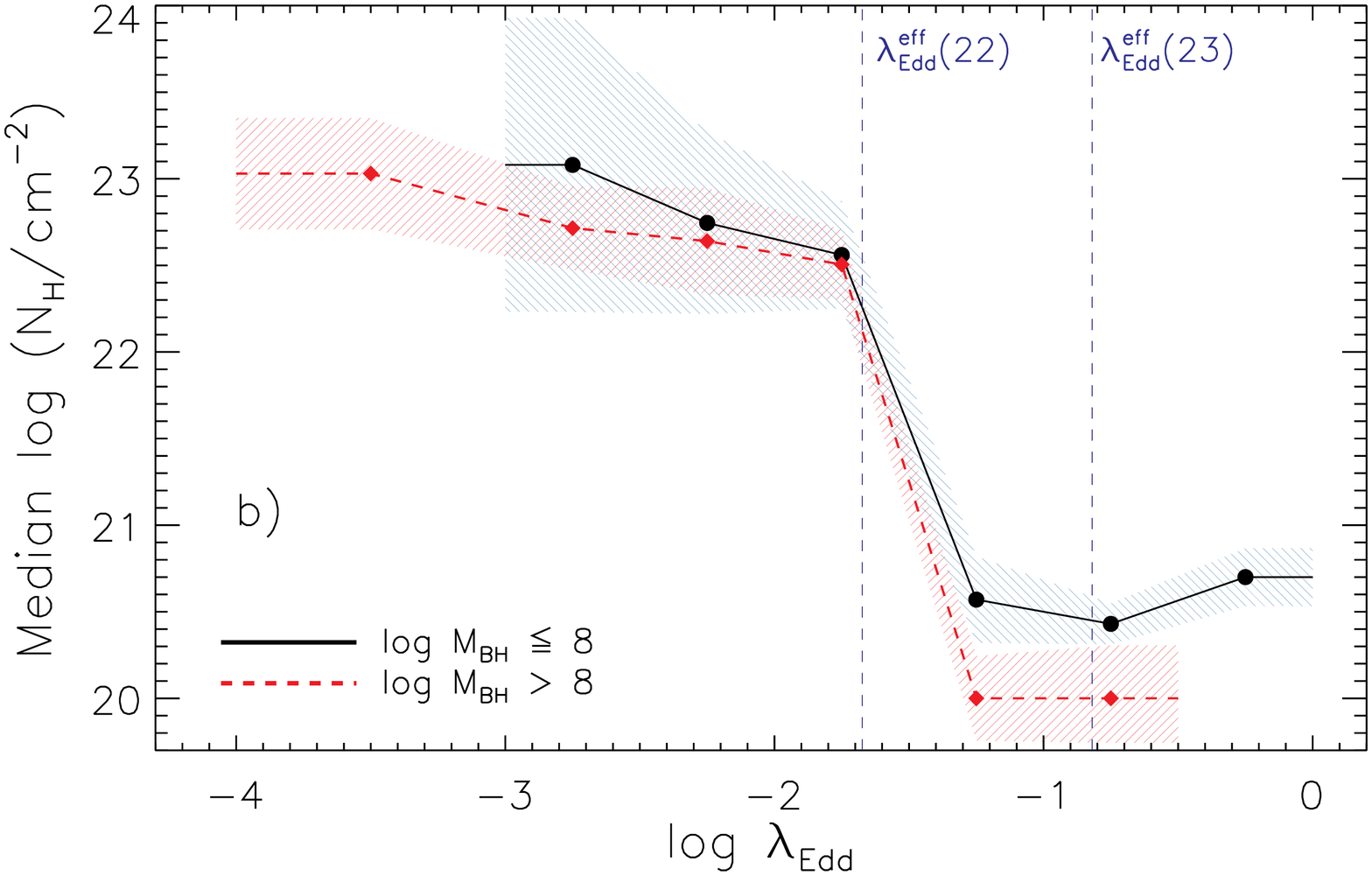}
 \caption{\textbf{Median value of the column density versus Eddington ratio for different luminosity and black hole mass ranges.} Same as Extended Data Figure\,\,\ref{fig:medianNH_vs_Eddratio} for two different ranges of the intrinsic 14--150\,keV luminosity (panel a, in $\rm\,erg\,s^{-1}$) and black hole mass (panel b, in $M_{\odot}$). The filled areas represent the median absolute deviations. The dashed vertical lines represent the effective Eddington limit for dusty gas with $N_{\rm H}=10^{22}\rm\,cm^{-2}$ [$\lambda_{\rm Edd}^{\rm eff}(22)$] and  $N_{\rm H}=10^{23}\rm\,cm^{-2}$ [$\lambda_{\rm Edd}^{\rm eff}(23)$] for standard dust grain composition of the interstellar medium. 
 }
 \label{fig:medianNH_vs_EddratioLumbins}
\end{figure*}

In agreement with the idea that radiation pressure is the main driver of nuclear obscuration, the median column density decreases very rapidly at $\lambda_{\rm Edd}\simeq 10^{-1.5}$ (with a significance of $\simeq 12\sigma$, Extended Data Figure\,\,\ref{fig:medianNH_vs_Eddratio}). A similar trend is observed dividing the sample into different ranges of the intrinsic 14--150\,keV luminosity and of black hole mass (panels a and b of Extended Data Fig.\,\ref{fig:medianNH_vs_EddratioLumbins}, respectively).

The radius of the sphere of influence\cite{Peebles:1972ts} ($r_{\rm SMBH}$) is $r_{\rm SMBH}=GM_{\rm BH}/\sigma_{*}^{2}$, where G is the Gravitational constant. Considering the relation between black hole mass and stellar velocity dispersion\cite{Kormendy:2013uf} [$M_{\rm BH}/10^{9}M_{\odot}\simeq 0.309\times(\sigma_{*}/200\rm\,km\,s^{-1})^{4.38}$] for black hole masses in the $M_{\rm BH}\simeq 10^6-10^9M_{\odot}$ range, the radius of the sphere of influence would be $r_{\rm SMBH}\simeq 1.5-63\rm\,pc$.

An increase in the fraction of obscured sources with the redshift has been found by several studies carried out in the past years\cite{Ueda:2014ix,Brightman:2014lq,Buchner:2015ve}. This could be possibly related to the observed increase in the amount of gas in galaxies with increasing redshifts\cite{Tacconi:2010ys}, which would straightforwardly lead to larger fractions of obscured sources. Moreover, both simulations and observations have shown that at higher redshifts galaxy mergers are also more common than in the local Universe\cite{Genel:2009rm,Rodriguez-Gomez:2015gf}. Considering the enhanced levels of obscuration observed in mergers, this could also contribute to the  increase of $f_{\rm obs}$ with redshift.

%*******************************************************************************************************************************************************************************

\section{The intrinsic column density distribution of sources accreting at different Eddington ratios}\label{sec:NHfunction}

While {\it Swift}/BAT observations are unbiased for $N_{\rm H}\lesssim 10^{24}\rm\,cm^{-2}$, they can still be affected by obscuration for $N_{\rm H}\geq 10^{24}\rm\,cm^{-2}$. We infer here the intrinsic $N_{\rm H}$ distribution of local accreting supermassive black holes. This was done by correcting the observed column density distribution for selection biases\cite{Marshall:1983eu,Ueda:2003qf,Ueda:2014ix} to obtain the ``$N_{\rm H}$ function'', i.e. the intrinsic probability distribution function of $N_{\rm H}$. For our {\it Swift}/BAT sample this was previously calculated in Ref.\,\citen{Ricci:2015tg}, where we also obtained the intrinsic $N_{\rm H}$ distribution by dividing the sample into two luminosity bins. Here we follow the same approach, separating sources into two Eddington ratio bins: $10^{-4} \leq \lambda_{\rm Edd} < 10^{-1.5}$ and $10^{-1.5} \leq \lambda_{\rm Edd} < 1$. This was done using the Eddington ratio obtained from the intrinsic 14--150\,keV {\it Swift}/BAT luminosities, using a bolometric correction of 8.47 (i.e. equivalent to $\kappa_{2-10}=20$ for $\Gamma=1.8$).
The intrinsic fractions of sources in different $N_{\rm H}$ intervals are listed in Extended Data Table\,\ref{tab:NH_eddratio}. The results show a very significant decrease in the fraction of Compton-thin obscured AGN for increasing Eddington ratio, with $f_{\rm obs}=64\pm5\%$ at $10^{-4} \leq  \lambda_{\rm Edd}< 10^{-1.5}$ and $f_{\rm obs}=19\pm4\%$ at $10^{-1.5} \leq  \lambda_{\rm Edd} < 1$. Normalizing the $N_{\rm H}$ function over the range $N_{\rm H}=10^{20}-10^{25}\rm\,cm^{-2}$, we find a fraction of Compton-thick AGN of $f_{\rm CT}=23\pm6\%$ for $10^{-4} \leq \lambda_{\rm Edd}< 10^{-1.5}$, and $f_{\rm CT}=22\pm6\%$ for $10^{-1.5} \leq  \lambda_{\rm Edd} < 1$.

\begin{table}[h!]
\caption[]{{\bf Intrinsic fraction of sources with a given column density in different ranges of $\lambda_{\rm Edd}$}. The table lists the (1) $N_{\rm H}$ range, and the fraction of sources within that  $N_{\rm H}$ range for (2) $-4 \leq \log \lambda_{\rm Edd}< -1.5$ and (3) $-1.5 \leq \log \lambda_{\rm Edd}< 0$. The values are normalized to unity in the $\log (N_{\rm H}/\rm\,cm^{-2})=20-25$ interval.}
\label{tab:NH_eddratio}
\begin{center}
\begin{tabular}{ccccc}
\hline
\hline
\noalign{\smallskip}
\multicolumn{1}{c}{(1)} & & (2) & & (3)  \\ 
\noalign{\smallskip}
$\log (N_{\rm H}/\rm cm^{-2})$&  & $-4$ to $-$1.5	&		& $-1.5$ to  0  \\ 
\noalign{\smallskip}
 & & {\footnotesize [\%]} &  & {\footnotesize [\%]} \\
\noalign{\smallskip}
\hline
\noalign{\smallskip}
\noalign{\smallskip}
20--21    &	 		&$8\pm6$			&	&		$45\pm4$			\\	%
\noalign{\smallskip}
21--22    &	 		&$5\pm1$			&	&		$14\pm2$			\\	%
\noalign{\smallskip}
22--23    &	 		&$26\pm3$			&	&		$7\pm3$			\\	%
\noalign{\smallskip}
23--24    &	 		&$38\pm4$			&	&		$12\pm3$			\\	%
\noalign{\smallskip}
24--25    &	 		&$23\pm6$			&	&		$22\pm7$			\\	%
\noalign{\smallskip}
\hline
\noalign{\smallskip}
\end{tabular}
\end{center}
\end{table}

\begin{addendum}
 \item[Data Availability Statement]  
The datasets generated during and/or analysed during the current study are available from the corresponding author on reasonable request. 
 \end{addendum}

\bigskip

\noindent {\Large \bf References for the Methods}

%\makeatletter
%\addtocounter{\@listctr}{30}
%\makeatother

\end{document}